# Natural Language Interfaces to Data

Abdul Quamar[1], Vasilis Efthymiou[2], Chuan Lei[3] and Fatma Özcan[4]

[1]*IBM Research AI, USA; ahquamar@us.ibm.com*
[2]*FORTH-ICS, Greece; vefthym@ics.forth.gr*
[3]*Instacart, USA; chuan.lei@instacart.com*
[4]*Systems Research@Google, USA; fozcan@google.com*

---

ABSTRACT

Recent advances in natural language understanding and
processing have resulted in renewed interest in natural lan-
guage interfaces to data, which provide an easy mechanism
for non-technical users to access and query the data. While
early systems evolved from keyword search and focused on
simple factual queries, the complexity of both the input
sentences as well as the generated SQL queries has evolved
over time. More recently, there has also been a lot of focus
on using conversational interfaces for data analytics, empow-
ering a line of business owners and non-technical users with
quick insights into the data. There are three main challenges
in natural language querying: (1) identifying the entities
involved in the user utterance, (2) connecting the differ-
ent entities in a meaningful way over the underlying data
source to interpret user intents, and finally (3) generating a
structured query in the form of SQL or SPARQL.

There are two main approaches in the literature for inter-
preting a user's natural language query. Rule-based systems
make use of semantic indices, ontologies, and knowledge

---

graphs to identify the entities in the query, understand the intended relationships between those entities, and utilize grammars to generate the target queries. With the advances in deep learning-based language models, there have been many text-to-SQL approaches that try to interpret the query holistically using deep learning models. Hybrid approaches that utilize both rule-based techniques as well as deep learning models are also emerging by combining the strengths of both approaches. Conversational interfaces are the next natural step to one-shot natural language querying by exploiting query context between multiple turns of conversation for disambiguation. In this monograph, we review the background technologies that are used in natural language interfaces, and survey the different approaches to natural language querying. We also describe conversational interfaces for data analytics and discuss several benchmarks used for natural language querying research and evaluation.

# 1

## Introduction

Natural language interfaces provide an easy way to query and interact with data, and enable non-technical users to investigate the data sets without the need for knowing a query language like SQL. As a result, natural language interfaces have been an active area of research for many decades. With the advances in natural language processing (NLP) technologies, and language models like BERT (Devlin *et al.*, 2019), there is renewed research interest. Even limited forms of such interfaces are now becoming available in commercial products (*Ask Data | Tableau Software* 2021; *Power BI Platform* 2021; *Cognos Assistant* 2021).

Many business users and line of business owners rely on technical people to query and gain insights from their data. These technical people are experts on using complex query languages such as SQL or SPARQL. Today, it is vital for non-technical uses to derive insights from their data as quickly as possible to make effective business decisions. Most often business owners do not have direct access to the data, instead relying on application interfaces with pre-defined queries or dashboards to access and examine the data. Usually, technical users close the gap by creating the dashboards and the canned queries needed, but this introduces delays. Today, there is an increasing need for rapid data access and insights as well as quick exploration of data as soon as it lands in the database. Natural language interfaces provide this functionality, giving rise to the augmented consumer (Richardson *et al.*, 2021). Gartner predicts that the future analytics experiences will be consumer-focused, augmented in context as well as conversational.

Natural language interfaces include natural language query (NLQ) systems, as well as dialogue (or conversational) systems. NLQ systems interpret a single user utterance and produce a SQL or SPARQL query. In other words, NLQ systems offer one-shot query answering, without any context between subsequent queries, whereas conversational systems allow multiple turns in question answering, while preserving some context between turns. This additional context information allows further disambiguation in interpretation.

There are several challenges in building natural language interfaces to data (Affolter *et al.*, 2019). Ambiguity in natural language is a big challenge, making it difficult to understand the semantics of the query and hence the user intent. Understanding the complex relationships between the entities in the user statement and generating a complex SQL query are also challenging. General purpose solutions that can be adapted quickly to any domain are difficult to build. Figure 1.1 shows the three important tasks that are involved in natural language querying of data. The first task in NLQ is semantic parsing and entity tagging, which identifies the entities involved in the user query. Identifying the relationships between these entities, associating them with the data elements in the database, and finally interpreting the user intent based on these entities and relationships is the most critical and challenging task in NLQ. There may be many interpretations that are valid and choosing the right one is also non-trivial. Finally, the last task in NLQ is generating the SQL query that corresponds to the chosen interpretation.

There are two main approaches to NLQ: rule-based and ML/DL-based techniques. Some systems (Saha *et al.*, 2016; Lei *et al.*, 2018; Sen *et al.*, 2019; Li and Jagadish, 2014b; Li and Jagadish, 2014a; Li and Jagadish, 2016; Blunschi *et al.*, 2012; Song *et al.*, 2015) use semantic indexes or ontologies to identify the entities in the query, and employ rule-based or grammar-based techniques for query interpretation and SQL generation. Machine learning (ML) and deep learning (DL) based text-to-SQL techniques (Basik *et al.*, 2018; Weir and Utama, 2019; Zhong *et al.*, 2017; Xu *et al.*, 2017; Yu *et al.*, 2018a; Gur *et al.*, 2018; Zhang *et al.*, 2019), which encode user inputs into a feature embedding and train deep learning models to generate the SQL query in a holistic way, are widely used, and have become more popular recently. While rule-based approaches provide easier domain adaptation, text-to-SQL systems are more robust to paraphrasing of the input query. There are also some emerging hybrid solutions that mix rule-based and ML/DL-based techniques for different NLQ tasks. For example, Usta *et al.* (2021) provide a DL-based technique for entity tagging that can be plugged in any rule-based solution.

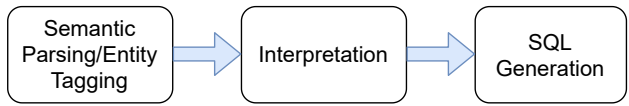

**Figure 1.1:** Tasks in natural language querying.

Natural language interfaces have been an area of active research in various communities for many years (Özcan *et al.*, 2020; Li and Rafiei, 2017; Affolter *et al.*, 2019; Katsogiannis-Meimarakis and Koutrika, 2021b; Gkini *et al.*, 2021). Figure 1.2 shows a historical timeline for many NLQ and conversational solutions. In particular, the search and NLP communities have worked on natural language interfaces by extending keyword search into templates and sentences. Many question answering systems are in this group. A question answering system allows the user to ask questions in natural language and to obtain direct answers that correspond to facts stored in the database. It can be considered as an enhancement to search systems. Instead of a simple keyword search over the data, question answering systems can provide more meaningful and insightful information in the form of short answers to the user's natural language questions. Similar to keyword search, the goal in these use cases is to find information about certain entities, such as the CEO of a company or the director of a movie. In these systems, the final structured query that gets generated is a simple lookup query. Examples include early systems (Aditya *et al.*, 2002; Tata and Lohman, 2008) that only allow a set of keywords, with very limited expressive power, as well as systems (Blunschi *et al.*, 2012; Zenz *et al.*, 2009) that mostly focus on simple queries that access a single table using some selection criteria. Later works allow a full-blown English statement and try to disambiguate among the multiple meanings of the words and their relationships. There has been also work on building conversational systems (Yu *et al.*, 2019a; Quamar *et al.*, 2020a) that allow advanced search on well-curated databases.

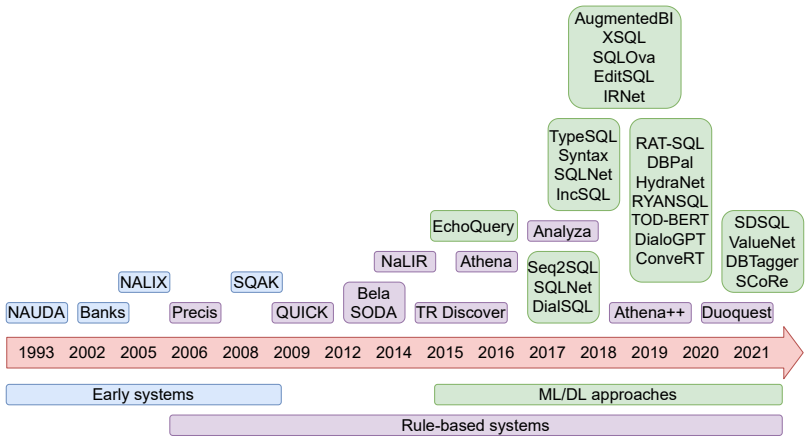

**Figure 1.2:** Historical perspective.

The database community has focused on natural language interfaces for analytical queries, as such interfaces enable business users and analytics teams to quickly analyze the data, and understand reasons and key drivers for business behaviors. As predicted by Gartner (Richardson *et al.*, 2021), to become more widely used than pre-defined dashboards, these systems require complex SQL queries that are typical in analytical systems. With the recent advances in NLP (Young *et al.*, 2018), both the complexity of input natural language statements, as well as the generated SQL and SPARQL queries have increased over time. A lot of these systems (Li *et al.*, 2005; Saha *et al.*, 2016; Sen *et al.*, 2020; Basik *et al.*, 2018) have originated in the database research community and can generate complex SQL queries with many joins and aggregations, as well as nesting.

In this monograph, we first review the background technologies empowering the existing natural language interfaces to data in Section 2. Then, in Section 3, we discuss many rule-based and text-to-SQL systems, as well as hybrid solutions to natural language querying. We also describe how to extend the one-shot query approaches to dialogue, taking advantage of the context for disambiguation, in Section 4. In Section 5, we recount various benchmarks designed for evaluating natural language interfaces to data. Finally, we conclude with a discussion on challenges that need to be addressed before these systems can be widely adopted in Section 6.

# 2

---

# Background

---

## 2.1  Data Modeling: Ontologies, Taxonomies, Knowledge Graphs

The rapid developments towards the Semantic Web vision in the past two decades have significantly impacted a plethora of fields in computer science, especially those targeting a deeper understanding and processing of data. Natural language interfaces to data could benefit from such developments. Having a formal semantic definition of the underlying data and making the data machine-readable and understandable were the two key factors towards providing a better understanding and a greater potential of answering user queries, expressed not only in a structured query language (e.g., SQL, SPARQL, Cypher), but also in natural language.

The key notion behind the Semantic Web is that of an *ontology*, which is typically based on the fragments of first-order logic called Description Logics (Baader *et al.*, 2017). An ontology is a formal conceptualization of a domain in terms of *classes*, representing entities, and *properties*, representing features of those entities, including their relationships to other entities. A typical example of an ontology is one that includes the classes 'Person', 'Professor', 'Student', 'Course', and the relationships 'teaches', associating a 'Professor' to a 'Course', and 'enrolls', associating a 'Student' to a 'Course' (Figure 2.1). Ontologies also contain subsumption (aka *is-a*) relations between classes (and even properties), e.g., every 'Professor' is-a 'Person'. In fact, ontologies that contain only subsumption relations are also referred to in the literature as *taxonomies*. Similar to database integrity constraints, ontologies may also contain some restrictions, such as domain and range restrictions for relationships (e.g., the domain of teaches is the class 'Professor', and its range is 'Course'), and functionality assertions (e.g., if the relationship 'teaches' is functional, this means that a Professor cannot teach more than one Course).

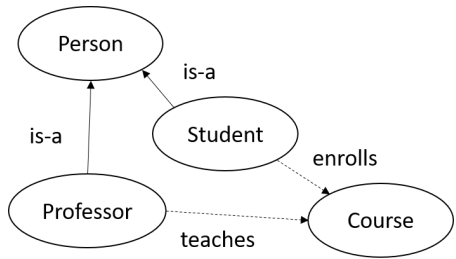

**Figure 2.1:** An ontology example.

By providing an ontology for the domain of a dataset, a computer system knows not only how to present the data, but also what the data is about (i.e., the meta-data), as well as how to process and even infer new information that is not explicitly stated in a given dataset (Özcan *et al.*, 2021). A typical inference example for the ontology of Figure 2.1 is that if Tom is an instance of a 'Professor', then we can infer that Tom is also a 'Person', even if this fact is not explicitly stated in the data. Those modeling and inference capabilities offered by ontologies have provided a great potential for rule-based natural language interfaces to data to bring new query answering capabilities, by associating chunks of natural language questions with formally defined ontological classes and properties.

A graph-structured representation of the meta-data (as provided by an ontology), as well as the data (aka facts) is commonly referred to as a *knowledge graph*. In this work, we will abstractly refer to knowledge graphs, regardless of the underlying data storage technology, which may be a relational database, an RDF triple store, or a graph database.

## 2.2 Language Models

Language models have a rich history of association with NLP (Qiu *et al.*, 2020). In this section, we provide a brief description of the evolution of language models and describe their usage and application for building natural language interfaces to data.

One of the earliest language models was to model a document as a bag of words for downstream classification tasks. TF-IDF (Term Frequency-Inverse Document Frequency) was used to normalize these word counts, enabling the scoring of words in ML algorithms for NLP tasks. TF-IDF allows the score to reflect the frequency of occurrence of a word, while offsetting it by the number of documents that contain the word, resulting in higher accuracy than the bag of words approach. This approach however does not capture the context of a word in a document with respect to its relationships with other words.

Word embeddings overcome this limitation by capturing the structure and relationships of words with respect to other words. Word2Vec (Mikolov *et al.*, 2013) is one of the most popular implementations of such a model that produces a vector representation (embedding) of a word, using shallow architectures, such as continuous bags of words and skip-gram models. These models learn embeddings using pairwise ranking, utilizing nearby words from the same sentence. They are pre-trained using a large corpus of textual data. Then, they are used to generate a latent vector representation of words such that the vectors of two semantically similar words are placed close together in a multi-dimensional latent space and vice-versa. GloVe (Pennington *et al.*, 2014) is another widely used model for obtaining pre-trained word embeddings based on a global word-word co-occurrence matrix constructed from a large corpus of textual data. The model is a count-based method that captures the proximity of words in terms of the ratio of their co-occurrence probabilities. These approaches generate word embeddings as vector representations in an unsupervised setting by using the raw corpus of textual data without any labels. These models can be used by downstream NLP classification tasks. However, they are context-independent, i.e., the embedding of a word is always the same irrespective of the context in which it is used. Also, these models are unable to handle out-of-vocabulary words, i.e., words that did not appear in the training data.

To distinguish the semantics of a word in the different contexts that it is used, what is needed is contextual embeddings. Models for generating contextual embeddings pre-train neural networks at the sentence and paragraph level. ELMo (Embeddings from Language Models) (Peters *et al.*, 2018) is a contextual embedding model that uses a 2-layer Long Short-Term Memory (LSTM) encoder with a bidirectional language model that takes into account the order of words before and after a particular word in both directions. ELMo embeddings have been shown to yield large improvements in a broad range of NLP tasks that require capturing word semantics based on their context and can also handle out-of-vocabulary words. Pre-trained language models such as ELMo can also be fine-tuned for downstream tasks (e.g., text classification) on different datasets, and have been shown to produce state-of-the-art results.

Transformers (Vaswani *et al.*, 2017) provide another way to capture long dependencies at the sentence and paragraph level. Transformers are based on an encoder-decoder architecture and use a bidirectional self-attention mechanism to encode the context of a given word with respect to the sentence in which it is used. They have been shown to capture long-term dependencies better than LSTM-based models.

BERT (Bidirectional Encoder Representation from Transformer) (Devlin *et al.*, 2019) is a transformer-based pre-trained language model that uses the encoder block of the transformer architecture and has shown to be very effective in learning universal language representations. The input to BERT is a sequence of token embeddings. The encoder consists of several stacks of bidirectional transformer encoder layers, and the output is a set of contextualized embeddings. BERT pre-trains deep bidirectional representations on a large corpus of data, such as Wikipedia (2.5B words) or BooksCorpus (800M words), while paying attention to context in both directions, in all layers.

Pre-training a BERT model can be done in two ways. First, a masked language model (Taylor, 1953) is used to randomly mask some of the tokens in the input with the goal of predicting the masked word based on its context (e.g., using "The capital of France is [MASK]." to predict

"Paris"). The masked language model allows fusing the context in both directions, hence enabling the pre-training of a deep bidirectional transformer. Second, next-sentence prediction can be used to learn the dependencies between different sentences. Given two sentences $S_1$ ("Home price growth is rapid") and $S_2$ ("It is really hard to buy a home in the Bay Area") as input, BERT is trained to predict if $S_2$ logically follows $S_1$, thereby learning the longer-term dependencies between sentences.

A pre-trained BERT model can be fine-tuned using additional layers and training to produce effective models shown to perform well for different downstream tasks. Each downstream task requires providing the task's input in the form of appropriate features to BERT and using BERT's output to make a prediction decision for the task. BERT has been shown to do well for several NLP tasks, such as semantic tagging and question answering. It has been also widely used in several text-to-SQL systems, such as HydraNet (Lyu *et al.*, 2020) and SQLova (Hwang *et al.*, 2019), which will be covered in detail in Section 3.2.

Since the widespread use of BERT, there have been many other variants, such as ALBERTA (Lan *et al.*, 2019), RoBERTa (Liu *et al.*, 2019), TinyBERT (Jiao *et al.*, 2020), DistilBERT (Sanh *et al.*, 2019), and SpanBERT (Joshi *et al.*, 2019), targeting different downstream applications. Since most language models are pre-trained on domain-agnostic corpora, their applicability to certain domain-specific tasks is limited. Language models pre-trained on domain-specific corpora have also been proposed recently, including BioBERT (Lee *et al.*, 2019) for biomedical text, SciBERT (Beltagy *et al.*, 2019) for scientific text, and FinBERT (Araci, 2019) for financial text.

Fine-tuning language models has now become a popular approach for adapting the pre-trained language models for several different downstream tasks. For example, ULMFit (Universal Language Model Fine-Tuning) (Howard and Ruder, 2018) particularly attempts to fine-tune models for text classification in two steps, namely (1) fine-tuning pre-trained language models on domain-specific or target data with the goal of domain adaptation, and (2) further fine-tuning language models for specific downstream tasks using techniques like discriminative fine-tuning. Discriminative fine-tuning uses different learning rates for different layers. This approach allows for partially retaining previous knowledge learnt from the initial training while adaptively learning new knowledge from the fine-tuning. This adaptive approach has been shown to work well empirically on different datasets for text classification.

A step further in fine-tuning language models is OpenAI GPT (Generative Pre-Training) (Radford and Narasimhan, 2018), which has demonstrated that a large number of NLP and NLU (natural language understanding) tasks, e.g., question answering and semantic similarity assessment, can benefit from generative pre-training of language models. GPT uses a 12-layer decoder block of the transformer architecture, wherein the language model is first pre-trained on a diverse corpus of unlabelled text. This is followed by discriminative fine-tuning on specific downstream tasks. Unlike other approaches discussed above, GPT affects transfer learning using task-aware input transformations during the fine-tuning stage, while the model itself is general and task-agnostic. GPT has been shown to perform well especially on question answering tasks such as RACE (Lai *et al.*, 2017), a benchmark dataset for reading comprehension tasks. GPT-2 (Radford *et al.*, 2019) further improves upon GPT and adds a normalization layer to the input of each sub self-attention block, as well as an additional layer of normalization after the final self-attention block. It was primarily designed as an auto-regressive language model for predicting the next word, given a sequence of prior words in a text. GPT-2 can have 24, 36 or 48 layers in the decoder block and has been trained on the WebText corpus, created by scraping web pages with emphasis on document quality.

Fine-tuning of language models, although being extensively used, still requires a substantial amount of task-specific or domain-specific labelled examples to achieve the desired goal. Scaling up language models can alleviate this issue by reducing the need for fine-tuning. The GPT-3 language model (Brown *et al.*, 2020) demonstrates this by obviating the need for any fine-tuning or gradient updates. GPT-3 has been shown to achieve good performance on several NLP tasks, such as question answering and translation, as well as tasks that require some reasoning or domain adaption, such as using new words in a sentence and handling arithmetic tasks. GPT-3 is characterized by its large model size (175 billion parameters) and has been tested in the few-shot setting, i.e., making predictions based on a limited number of samples.

Recently, newer language models such as Transformer-XL (Dai *et al.*, 2019) and XLNet (Yang *et al.*, 2019) have been proposed to overcome some of the limitations of transformer-based language models, such as fixed-length context, context fragmentation, and neglecting dependency between masked positions. Transformer-XL enables learning beyond a fixed-length context by reusing hidden states from previous segments to build a recurrent connection between segments. XLNet, a generalized auto-regressive pre-training method, overcomes the limitations of BERT by enabling learning of bidirectional contexts through maximizing the expected likelihood over all permutations of the factorization order.

Current language models are pre-trained on free-form text corpora and hence are limited in their ability to represent the two-way structure or multi-turn dynamics of a dialogue or conversation. Conversational semantic parsing tasks require the translation of a sequence of natural language queries to structured queries, such as SQL or SPARQL. SCoRe (Yu *et al.*, 2021) addresses the limitations of the current language models for conversational semantic parsing tasks by adapting pre-trained language models using a second phase of pre-training that captures both the multi-turn dynamics, as well as the structural contexts in a dialogue. LaMDA (Language Models for Dialogue Applications) (Google, 2021c) is a transformer-based language model trained on dialogues based on the fact that words in a dialogue across multiple statements are related and together they are sensible.

Other language models that have been used for tasks such as dialogue response generation include DialoGPT (Zhang *et al.*, 2020a) and ConveRT (Henderson *et al.*, 2020). DialoGPT is a dialogue Generative Pre-Trained Transformer trained on conversational data from Reddit and has been shown to generate relevant and context-sensitive responses to user queries in a conversational setting. ConveRT (Conversational Representations from Transformers) is a pre-training framework for response generation suitable for conversational settings, owing to its reduced model size and faster training time as compared to standard sentence encoders.

For more details on pre-trained language models for natural language processing, we refer the reader to (Qiu *et al.*, 2020). Next, we will describe how language models have been extensively used for NLP tasks relevant to natural language interfaces to data, such as semantic tagging, named entity recognition, natural language generation, and conversational systems.

## 2.3 Semantic Tagging and Named Entity Recognition

Semantic tagging, which has extensive applications in text mining, predicts whether a given piece of text conveys the meaning of a given semantic tag (Abzianidze and Bos, 2017). Named entity recognition (NER) relies on tagging words or phrases with semantically informative tags (e.g., person, location, organization, time) in text. In natural language interfaces to data, semantic tagging/NER is heavily used by both traditional rule-based and recent ML-based approaches. The goal of semantic tagging/NER in these systems is to annotate tokens in a text with the corresponding semantic information from the data.

There are two types of methods for semantic tagging/NER: rule programming (Li *et al.*, 2020) and machine learning (Bjerva *et al.*, 2016). Early systems (Sekine and Nobata, 2004; Negi and Buitelaar, 2015) require domain experts to handcraft rules and extract features for semantic tagging. These rules and features are often based on predefined dictionaries as well as the word- or document-level characteristics from the corpus. Such rule-based methodology is often error-prone and requires significant programming effort.

In contrast, modern semantic tagging/NER solutions (Kim *et al.*, 2016; Santos and Guimarães, 2015; Huang *et al.*, 2015; Shao *et al.*, 2016; Zhang *et al.*, 2020b) do not require much programming effort as they most often resort to machine learning techniques, deep learning models in particular. The primary reason is that they are often more capable of learning complicated functions than other kinds of models. Some prevalent deep models are based on convolution neural networks (Kim, 2014), LSTM (Hochreiter and Schmidhuber, 1997), and BERT (Devlin *et al.*, 2019).

Convolutional Neural Networks (CNNs)(Albawi *et al.*, 2017) tokenize a text into unigram words and each word is represented with a pretrained *k*-dimensional vector. A semantic tagging/NER model (Kim *et al.*, 2016) often uses highway networks over CNNs on character sequences of words and then uses another layer of softmax for the final predictions. Similarly, a deep learning model (Santos and Guimarães, 2015) is designed with a CNN over the characters of words, concatenated with word embeddings of the central word and its neighbors, fed to a feed-forward network, and followed by the Viterbi algorithm (Forney, 1973) to predict labels for each word.

LSTMs(Sherstinsky, 2020), based on Recurrent Neural Networks (RNNs), use the same input representation as CNNs, i.e., a matrix of word vectors. However, unlike CNNs, LSTMs sequentially (left to right) process the text over time and keep their hidden state through time. The hidden state can capture any meaningful features that appeared in the prefix of the text up to the current timestamp. This enables LSTMs to capture arbitrary long-term dependencies. With slight variations, bidirectional LSTM (bi-LSTM) and window bi-LSTM can be used to improve the performance (Huang *et al.*, 2015). Adding features, such as conditional random fields, parts-of-speech tagging, and case information, has also been shown to improve performance (Shao *et al.*, 2016).

As described in Section 2.2, BERT also uses a matrix of word vectors to represent a text, which is similar to CNN and LSTM. BERT applies attentions to represent a text with weighted word vectors, such that relevant tokens have higher weights than irrelevant ones. It has been shown that models using BERT as token-level embeddings for semantic tagging (Figure 2.2) are effective and robust over a variety of datasets (He and Choi, 2019). Similarly, SemBERT (Zhang *et al.*, 2020b) introduces an improved language representation model based on BERT, which absorbs contextual semantics from BERT and fine-tunes it without substantial task-specific modifications.

## 2.4 Natural Language Generation

Natural Language Generation (NLG) is the generation of human-readable text from a non-linguistic representation of information (Gatt and Krahmer, 2018). Non-linguistic inputs often include a variety of sources, such as semantic representations of information in different formats (e.g., JSON, OWL), data tuples coming from a relational database, information from structured knowledge graphs, data visualizations etc.

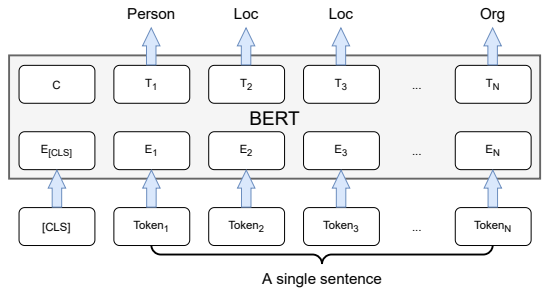

**Figure 2.2:** Semantic tagging using BERT.

More specifically, in the context of natural language interfaces to data, NLG is required to provide a human-readable text response that is typically accompanied with the results to a user query against the data. This human-readable text provides a summarized explanation of the accompanying results. For example, in response to a user query "Show me the sales by region for 2020", a natural language response accompanying the results could be: "Here are the sales by region for the year 2020". Techniques for NLG could be applied to both one-shot query answering systems as well as multi-turn conversational systems for data exploration.

A typical NLG task can be broken down into several components (Reiter and Dale, 1997; Reiter and Dale, 2000), which broadly determine the information content (what needs to be conveyed) and linguistic articulation (how to convey). These are (1) *content determination* to ascertain different pieces of information that need to be included in the generated text, (2) *text structuring* to determine the order of different pieces of information together, (3) *sentence aggregation* to determine which pieces of information would go in which individual sentences, (4) *lexicalization* to find the right words, phrases to express different pieces of information, (5) *referring expression generation* to determine words and phrases to represent entities or domain objects, and finally (6) *linguistic realization* to combine all constituent words and phrases into coherent text/sentences.

There are several different NLG architectures and approaches that have been used in the recent past to accomplish the aforementioned

NLG tasks. These range from (1) *modular architectures* (Reiter and Dale, 2000; Banaee *et al.*, 2013), which divide each of the sub tasks mentioned above into well-defined sub-modules and then stitch them together in the right order, (2) *planning perspectives* (Rieser and Lemon, 2016; Dethlefs, 2014; Garoufi, 2014), which view the process of NLG as a planning task allowing for a more integrated approach as compared to a modular architecture, and (3) *integrated or global approaches*, which form the most pre-dominant NLG technique that rely on statistical learning (Mairesse and Young, 2014) of the mapping or correspondences between the non-linguistic inputs to the text output of NLG often taking a holistic or global view of the NLG process. Each of these different architectures can potentially incorporate a wide variety of knowledge-based (symbolic) methods or data-driven stochastic methods to accomplish the different sub-tasks for NLG (Gatt and Krahmer, 2018). In this monograph, we focus on deep learning methods, encoder-decoder architectures and conditioned language models as the dominant stochastic methods used by the integrated approaches to NLG for building natural language interfaces to data.

Deep neural network architectures are well suited for sequence-to-sequence (Seq2Seq) translation and hence can be used for NLG tasks such as generating texts from abstract representations of information, a typical NLG requirement for natural language interfaces to data. Encoder-decoder architectures allow encoding the input sequence into vector representations which can then be decoded to support Seq2Seq tasks such as machine translation wherein a variable-length input sequence can be translated to a variable-length output sequence (Bahdanau *et al.*, 2015). Castro Ferreira *et al.* (2017) adapt the encoder-decoder architecture for Seq2Seq models to generate texts from Abstract Meaning Representations. Attention-based mechanisms further improve machine translation tasks by allowing the model to learn different weights for different parts of an input encoding, while decoding for different parts of an output (Bahdanau *et al.*, 2015). Such attention mechanisms thus allow a data-driven approach to learn correspondences between different parts of the abstract representations of information, such as a SQL query to different parts of natural language sentences. The attention mechanisms eliminate the need for a direct alignment between the input and output reducing the chances of error and making the translation more robust. This is particularly relevant for response generation for natural language interfaces to data.

Another way of generating natural language responses is to use a conditioned language model wherein the response is generated by sub-selecting portions of the input from a distribution conditioned on the input features. These features could be chosen based on the semantic, contextual or linguistic information content associated with them and the model jointly optimizes for multiple NLG tasks such as content selection and linguistic realization. Application of these conditioned models allows for generation of natural language text such as biography sentences from Wikipedia tables or infoboxes (Lebret *et al.*, 2016). The ability of conditioned language models to selectively incorporate specific contexts (both categorical and quantifiable or continuous attributes) to generate a summarizing text (Herzig *et al.*, 2017; Asghar *et al.*, 2017) from data originating from tables makes them attractive for generating natural language responses accompanying structured results for natural language interfaces such as Q&A and conversational systems. Wiseman *et al.* (2021) propose another approach for data-to-text generation using a combination of retrieval and generative methods. Candidate text(s) from a database are extracted based on a user query. The retrieved text(s) and the query are then combined in an utterance generator model to create a natural language response. A splicing technique is proposed to improve interpretability and provenance tracking. Such approach makes it clear which pieces of text are derived from which pieces of information from the database and how different pieces of text are stitched together to form a natural language response.

Extending NLG for response generation in conversational interfaces to data requires capturing context over multiple turns of conversation. This context is used as input for the encoder, while the decoder is used to generate the next dialogue response in natural language. Wen *et al.* (2015) propose an NLG technique for generating the next dialogue response using a semantically conditioned LSTM that learns from unaligned data by jointly optimizing sentence planning and surface realization. Several Seq2Seq models with attention mechanisms have also been used for generating the next response in conversational systems (Dusek and Jurcicek, 2016). As mentioned in Section 2.2, language models such as DialoGPT (Zhang *et al.*, 2020a) and ConveRT (Henderson *et al.*, 2020) have also been used for response generation in conversational systems.

For more details on NLG techniques and systems, we refer readers to Gatt and Krahmer (2018) and Santhanam and Shaikh (2019), which provide more in-depth descriptions and analysis of NLG techniques relevant to a variety of different systems including question-answering and dialogue systems.

## 2.5   Conversational Systems

Conversational systems are becoming increasingly popular as the preferred natural language interface to data exploration and analysis. These systems enable interaction with the framework through a dialogue, using multi-modal inputs including text, audio and gestures like touch or point. The ability of conversational systems to persist context across multiple turns of the dialogue allows users to interact and conduct a conversation with the frameworks, making them much more useful than one-shot query answering systems. Hussain *et al.* (2019) provide an in-depth survey on the history and evolution of conversational systems and their design techniques.

The prolific growth of conversational systems in different application domains has led to the development of a variety of conversational systems (or chatbots) to address different use cases. These systems can be broadly classified using two different criteria (Figure 2.3). (1) Based on the task (end goal) or user application: *task-oriented* and *non-task-oriented* chatbots. Task-oriented chatbots are useful for accomplishing specific tasks, such as making a travel reservation, executing a banking transaction, etc. Non-task-oriented chatbots are general-purpose, mostly used as information retrieval systems not tied to any particular task. (2) Based on the domain of data and knowledge: *open-domain* or *domain-specific* chatbots. As the name suggests, open-domain chatbots are not tied to any particular domain. These open-domain agents, such as Microsoft's Cortana (Microsoft, 2018), Apple's Siri (Apple, 2018), Google Assistant (Google, 2021b) and Amazon's Alexa (Amazon, 2018), are useful for accomplishing simple day-to-day tasks, such as checking weather forecasts, playing music, setting device timers, and general-purpose information retrieval. Open-domain chatbots are typically designed with a general-purpose front-end, which receives user input and, depending on the type of question asked, hands it to one of several backend services designed to handle the specific question. On the other hand, domain-specific chatbots are designed with a specific domain (e.g., finance, healthcare, transportation, insurance) in mind and hence, are more useful for in-depth conversational interactions for these application domains. A more detailed classification for conversational systems can be found at (Hussain *et al.*, 2019). Conversational interfaces to data that allow users to explore and analyze data using natural language over multiple turns of conversation often fall under this categorization depending on the kind of data being analyzed/explored.

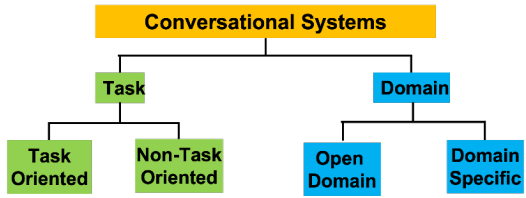

**Figure 2.3:** Classification of conversational systems.

Building conversational systems for different use cases and applications is facilitated today by the availability of several cloud-based chatbot platforms (e.g., Google Dialogflow, Facebook Wit.ai, Microsoft Bot Framework, IBM Watson Assistant, etc.). Using these platforms, developers can create many kinds of conversational systems for a variety of domains (e.g., weather, music, finance, travel, healthcare). Each of these platforms enables the design of different components (or building blocks) for a conversational system and customizes them for different use cases, applications and domains that the user is interested in.

The space of all possible user interactions (Figure 2.4) with a conversational system is defined in terms of three main components: *intents*, *entities*, and *dialogue* (Özcan *et al.*, 2020; Gao *et al.*, 2018). Intents express the purpose/goal or specific intended action as discerned from the user utterances. Entities represent information that is relevant to the user's purpose. These entities would typically consist of elements or real-world objects from the domain schema, as well as actual data instances that are relevant to conversation or the context of the user query. Entities therefore represent the vocabulary supported by the conversational system. The identification of intent and entities thus constitutes the natural language understanding component of the system.

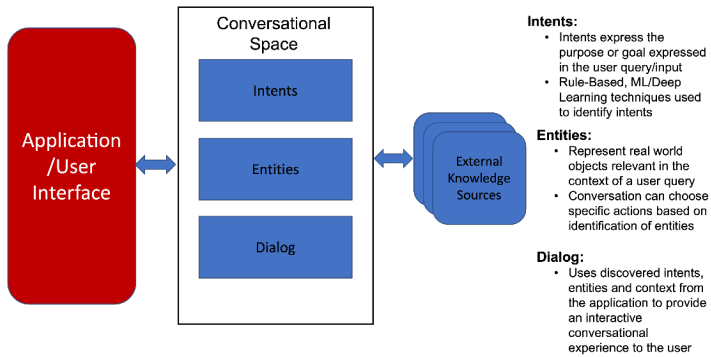

**Figure 2.4:** Conversational system: basic building blocks.

Dialogue defines interaction patterns supported by the conversational system. More specifically, it has three primary tasks: (1) *dialogue state tracking*, (2) *decision making*, and (3) *generating the natural language response* (Gao *et al.*, 2019). The dialogue subsystem provides a natural language response to a user conditioned on the identified intents, extracted entities in the user's input and the current context of the conversation persisted across multiple turns of the conversation using dialogue state tracking (history of conversation). Together, the intents, entities and dialogue form the basic building blocks of any conversation system irrespective of its type and application domain. Figure 2.5 shows an example workflow of a conversational system wherein a user requests for a set of specific movies, the natural language understanding component of the system identifies the intent as a movie request, the relevant entities mentioned and produces a structured semantic representation. The dialogue takes this as input, updates the conversational context and takes action to get the information from a movie database using a structured query. The results of the query are conveyed to the user along with a natural language response.

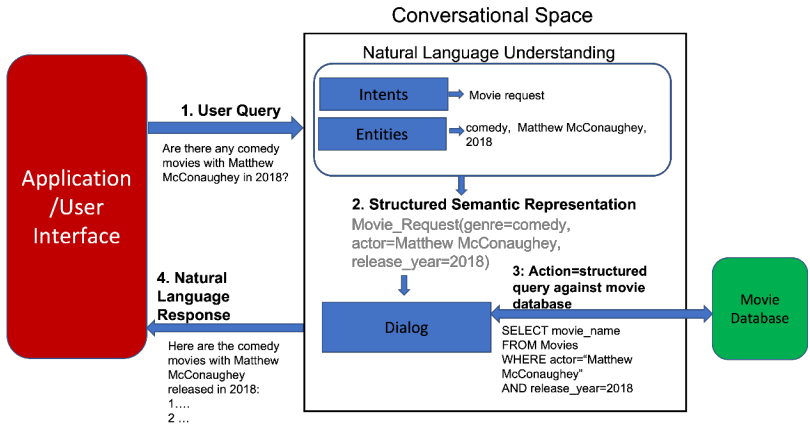

**Figure 2.5:** Example workflow of a conversational system.

Similar to other natural language interfaces, the identification of intent and entities from user utterances in a conversational interface requires parsing of user utterances, entity extraction and recognition (described in Section 2.3), encoding of the parsed input and its interpretation to identify user intents (Katsogiannis-Meimarakis and Koutrika, 2021a). What distinguishes intent identification in conversational systems is the inclusion of the conversational context persisted across several turns of prior conversation while interpreting the current user utterance. Depending on the identified intent and entities, the dialogue generates natural language responses using NLG techniques described in Section 2.4. These natural language responses are often accompanied by actual information or results retrieved from a variety of external systems such as knowledge graphs, relational databases, Business Intelligence (BI) platforms, search engines, or other information retrieval systems. Such information retrieval from external systems requires generation of an appropriate structured query or API call based on the user's question or utterance. The NLG response generated by the dialogue in such cases is thus conditioned on the user's natural language query (the identified intent and entities), the conversational context (the in-formation persisted from prior queries), the resulting action in terms of a structured query executed against an external data source and the results retrieved.

There are several techniques and approaches that are used for accomplishing each of the steps mentioned above, as well as for enabling conversational systems with domain-specific knowledge and understanding. These include rule-based approaches for semantic parsing and interpretation (Saha *et al.*, 2016), ontology-driven systems (Quamar *et al.*, 2020b) and various deep learning techniques using pre-trained and fine-tuned language models for both intent identification, structured query generation and natural language response generation. We discuss each of these in detail in Section 4, particularly with regard to conversational interfaces for data analysis and exploration.

**Summary**

In this section, we provide a brief overview of different techniques employed by natural language interfaces to data. We describe relevant data modeling techniques such as ontologies, taxonomies and knowledge graphs that help in formalizing the semantics of the underlying data. This enables better interpretation of the user queries in natural language and identification of relevant data to respond with. We describe the evolution of language models that have emerged as a powerful tool in understanding the semantics of natural language utterances and providing succinct low dimensional representations useful for several downstream tasks, including semantic tagging and named entity recognition, which are essential for building natural language interfaces to data. This section also covers techniques necessary for generating natural language responses to user queries that are often accompanied with the results retrieved from the data sources. Finally, we provide an overview of conversational systems including a classification of such systems, their basic building blocks and functionalities.

# 3

## Natural Language Querying Architectures

At the core of natural language querying (NLQ) systems lies their abilities to understand/interpret a user query expressed in natural language and to generate a structured query to be executed against a structured data source. The main challenges in building such a natural language querying (NLQ) system can be broadly categorized into the following three areas.

*Natural language understanding and query interpretation.* The inherent characteristics of natural language queries makes the query understanding and interpretation difficult. These characteristics often include ambiguity in terms of intent and entities expressed in the query, implied query context, linguistic variations, and partial or incomplete queries such as those that are expressed as keyword searches. An NLQ system is required to discern the context of such queries to provide appropriate answers.

*Domain understanding and generalizability.* Different domains, such as finance and healthcare, have their own unique characteristics and vocabulary. An NLQ solution should not only understand the semantics of a particular domain, but also be adaptable across different domains.

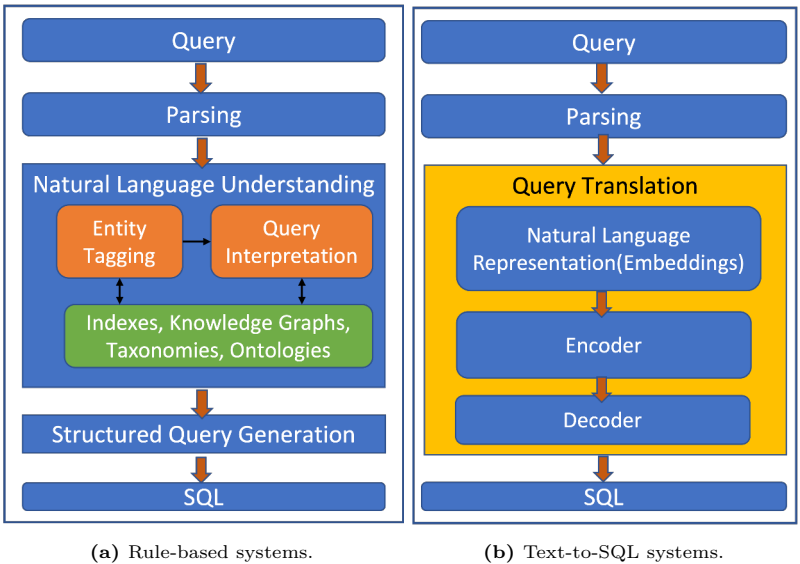

**(a)** Rule-based systems.        **(b)** Text-to-SQL systems.

**Figure 3.1:** Natural language querying architectures.

*Structured query generation.* The complexities of the structured queries like SQL and SPARQL make the query translation from natural language very challenging. An NLQ system needs to infer appropriate entity mappings from a natural language query to schema elements and to derive correct query structures from linguistic patterns embedded in a query.

There are numerous NLQ systems and architectures that attempt to address the aforementioned challenges. They can be classified based on their interpretation and structured query generation methods. Figure 3.1 shows a broad overview of NLQ system architectures. It captures two prevalent paradigms in the field, rule-based systems and ML/DL-based text-to-SQL systems. Both of these architectures require an input query parsing layer, which involves parsing a natural language query and transforming it into a more structured intermediate representation.

Following parsing, the rule-based paradigm takes a two-step approach. The first step involves identifying the entities in the user query, referred to as semantic entity tagging or entity extraction, and generating different interpretations based on an underlying knowledge model (e.g., indexes, knowledge graphs, ontologies). The second step is structured query generation, which leverages the mappings between the entities in the natural language query and the underlying schema elements and translates the query interpretations into a structured query like SQL that can be executed against a specific schema.

The other paradigm is the ML/DL-based approaches commonly referred to as *text-to-SQL* approaches, where trained models take texts as input and generate a structured query like SQL in a holistic way. Figure 3.1 shows a broad overview of different components in these text-to-SQL systems. Typically, a text-to-SQL system includes the following high-level steps: (1) natural language representation that transforms the entities in the input query text into feature vectors or embeddings using different techniques such as one-hot encoding or pre-trained language models, (2) encoder, a neural network that takes as input the feature vectors of the query text and the schema elements, and finally (3) decoder, that decodes the learnt intermediate representation provided by the encoder to produce the final SQL query.

Each of these architectures have their strengths and limitations affecting their adoption for different use cases. We describe these in further detail in the next two sections. There also exist a few hybrid approaches, an active area of research, that leverage the strengths of both rule-based and text-to-SQL approaches to build effective natural language interfaces to data.

## 3.1 Rule-Based Approaches

Rule-based approaches identify the entities mentioned within a query, as well as the relationships between entities, based on an internal or external representation (e.g., an inverted index, a taxonomy, or an ontology) of the underlying data. To tackle the semantic tagging challenge (described in Section 2.3), a typical rule-based system uses an external NLP library (e.g., Stanford's CoreNLP), or assumes that the input query is simple to process (e.g., a (set of) keyword(s)), or utilize certain formatting rules. This makes entity tagging much easier, allowing those rule-based systems to focus on the more involved task of query interpretation. Query interpretation requires understanding of the semantics of the identified entities and their relationships. It often requires references to indexes, external knowledge graphs and ontologies.

Once the natural language understanding (entity tagging and query interpretation) is finished, structured query generation can take the resulting query interpretation(s) and translate them into the target query language, again following some translation rules. For example, different query keywords (like "HOW MANY", "BY YEAR") may be associated with different query operations (like "COUNT()", "GROUP BY [YEAR]", respectively). More complex rules that generate nested queries are also applicable in this approach, as described next.

We start this section with a deeper dive into the components of one of the latest such approaches (Sen *et al.*, 2020) that uses ontologies for meta-data representation, and then discuss earlier approaches.

ATHENA (Saha *et al.*, 2016; Lei *et al.*, 2018; Sen *et al.*, 2019) and ATHENA++ (Sen *et al.*, 2020) associate parts of a given natural language query to concepts and relationships in an ontology that models the semantics of the data stored in a relational database. An initial ontology is generated automatically by ATHENA from the underlying database (Jammi *et al.*, 2018) that captures information available in the database schema such as tables, columns and their data types, primary key - foreign key constraints, etc. This basic ontology is further enriched with semantic information learnt from richer external reference ontologies such as FIBO, SNOMED, using techniques described in (Hao *et al.*, 2021). ATHENA and ATHENA++ use Ontology Query Language (OQL) as an intermediate query language that provides an intermediate abstract representation of the query against the ontology and helps in translating the input natural language query into SQL.

In the example of Figure 3.2, the input sentence "Show everyone who sold bonds in 2021 that have gone down in value" is first parsed to associate tokens (single words or multiple words) to candidate elements from the underlying data, which have been modeled as an ontology. For example, the token "bonds" is associated with the ontology element "ListedSecurity". The concept "ListedSecurity" in the ontology maps to a table with the same name in the underlying relational database. In a similar way, ATHENA assigns properties to the classes of the ontology, like the property "value" for the class "MonetaryAmount", as well as infers restrictions for those properties (e.g., functionality, inheritance, etc), which further help the query interpretation.

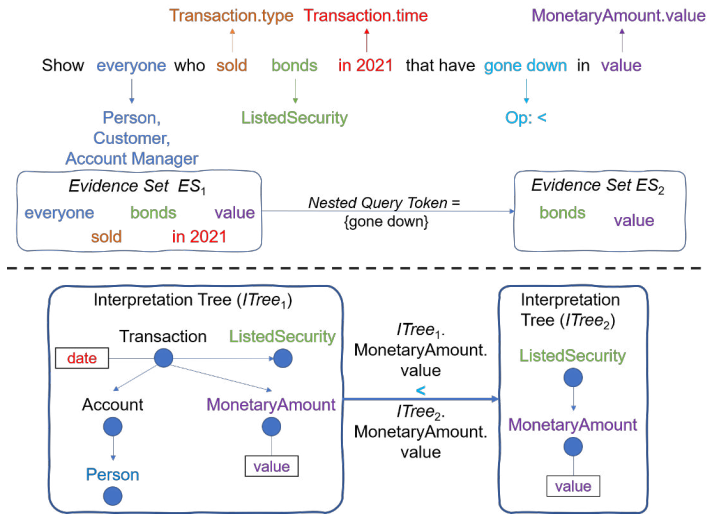

**Figure 3.2:** ATHENA++ example (Sen *et al.*, 2020).

The parsing of tokens in a given sentence is performed through common NLP libraries, like Stanford's CoreNLP, as well as rule-based annotators for person and time references (e.g., "everyone" typically refers to a person, "in 2021" is a year reference which is captured in the ontology as "Transaction.time"). The candidate entities from the ontology, which can be more than one per token (e.g., see the candidate ontology elements for the token "everyone") form the Evidence Set, and they are then connected to each other in the ontology graph, forming an interpretation tree (Figure 3.2, bottom). A Steiner Tree-based algorithm (Saha *et al.*, 2016), i.e., a generalization of the minimum spanning tree problem, is used to identify the most coherent interpretation. In addition, ATHENA++ (Sen *et al.*, 2020) has a nested query detection mechanism, which can detect hints for nested queries, such as "gone down" in our running example. This splits the given sentence in two (or more) sub-queries, each having a separate interpretation tree. The two trees are connected via the token "value", which has been found to correspond to the ontology element "MonetaryAmount.value". The "value" token is used as the join condition between inner and outer queries.

As a last step, ATHENA and ATHENA++ translate the interpretation tree(s) into an intermediate query language, which can be then seamlessly translated into the desired query language. The use of an intermediate query language decouples the query interpretation phase from the actual data store used. That's because using exactly the same intermediate query language as input, different structured query generators (a.k.a translators) can be created to target a different query language, e.g., one translating the intermediate query into SQL and another one translating the intermediate query into SPARQL.

To further exploit the semantics of the ontology to improve the query understanding capability of ATHENA, Lei *et al.* (2020) introduce a query relaxation technique by leveraging external knowledge sources, with a focus on medical KBs. This query relaxation method aims to fill the gap between the terms stored in the medical KBs and the colloquial or imprecise terminology in user queries. More recently, Ahmetaj *et al.* (2021) further allow the computation of ontology-enriched query answers by leveraging the chase procedure from the data exchange community (Fagin *et al.*, 2005) using an external reference ontology. This enables to not only return additional answers to user queries over the underlying medical KB, but also to understand and answer new queries, which would not be meaningful without the external reference ontology.

Next, we describe rule-based systems which address the challenges in natural language understanding and structured query generation using different techniques, categorized based on whether those techniques rely on inverted indices, taxonomies, or ontologies and knowledge graphs. When possible, we draw parallels to the different components of ATHENA and/or the general architecture shown in Figure 3.1 (a).

### 3.1.1   Index-based Systems

Among the earliest rule-based approaches are Précis (Koutrika *et al.*, 2006; Simitsis *et al.*, 2008) and QUICK (Zenz *et al.*, 2009), which first parse the natural language query to make it machine-readable (what we previously described as semantic tagging and NER), and then look up the query sub-strings that correspond to entities, using a simple index structure. For example, Précis (Koutrika *et al.*, 2006; Simitsis *et al.*, 2008) converts an input query, consisting of a set of keywords, to disjunctive normal form (e.g., ["Stanley Kubrick" AND "drama"] OR ["Woody Allen" AND NOT "comedy"]) and then seeks candidate interpretations for each query disjunct in the underlying database using an inverted index of names (e.g., it looks for database records containing "Stanley Kubrick" and "drama", as well as for records containing "Woody Allen" but not "comedy"). QUICK (Zenz *et al.*, 2009) builds an inverted index on the entity instances of the underlying data, and for each query (given as keywords), it tries to bind its keywords to the elements of the inverted index. In addition, users can interact with the results returned by QUICK and select the query interpretation that best fits their query, among the suggested options.

Even if it is not a traditional index-based system, one of the most recent works that leverages an index is Duoquest (Baik *et al.*, 2020), which efficiently explores the search space for possible queries from a given natural language query, by employing a guided partial query enumeration. To help an index-based NLQ system select the best fitting structured query from its pool of candidate query interpretations, users can also provide some examples of the answers they are looking for (following a programming-by-example approach), in the form of a so-called Table Sketch Query. Duoquest can then easily eliminate candidate query interpretations that would not produce the answers the user provided.

### 3.1.2 Taxonomy-based Systems

To better capture the semantics of the underlying data, more recent works exploit richer semantic models than simple inverted indices, such as taxonomies. NaLIR (Li and Jagadish, 2014b; Li and Jagadish, 2014a; Li and Jagadish, 2016) uses Stanford's NLP Parser (Marneffe *et al.*, 2006) for semantic tagging, this way obtaining a linguistic understanding of a given query, represented as a parse tree (Figure 3.3(a)). The nodes of the parse tree correspond to entities that are associated to entities in the underlying data using the WUP (Wu and Palmer, 1994) WordNet-based similarity function. This may result in multiple entity associations per tree node (e.g., "VLDB" node in Figure 3.3(a) may refer to the VLDB conference, the VLDB Journal, or PVLDB), which are then clarified by users. After this user-assisted disambiguation and some further processing (e.g., checks for grammar validity, node proximity in original schema), a final refined parse tree (Figure 3.3(b)) is generated and then translated into SQL.

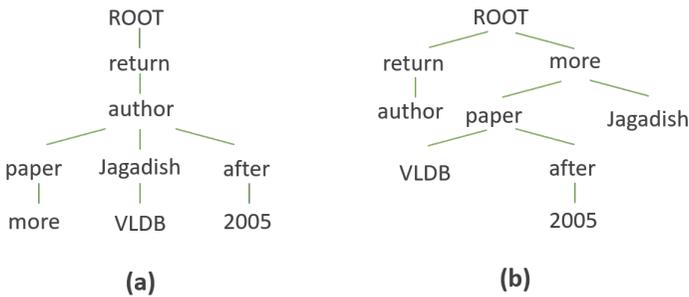

**Figure 3.3:** NaLIR's internal representation for the query *"show all authors who have more papers than H. V. Jagadish in VLDB after 2005"*, adapted from (Li and Jagadish, 2014b). (a) The initial parse tree from Stanford NLP. (b) A refined parse tree after some processing and user interaction for disambiguation.

### 3.1.3   Knowledge Graph/Ontology-based Systems

Even though taxonomies are clearly more expressive than simple inverted indices, there are still semantics involved in a query and in an underlying data source, which taxonomies cannot capture, the most prominent of which is relationships of different types between entities. Such semantics, as well as more conceptual modeling capabilities (e.g., functionality, symmetry, transitivity, role hierarchy) can be better captured in ontologies. SODA (Blunschi *et al.*, 2012) is among the first works to employ ontologies for query interpretation. In addition to an inverted index that it uses to look up query keywords in the data, it also looks up each keyword in an ontology, using a second meta-data-level index. For example, for a keyword query "clients Amsterdam assets", the keyword "clients" is found as a class in the underlying ontology, the keyword "Amsterdam" is found in the data, under the values of an attribute "addresses.city", while the keyword "assets" is found both in the ontology and in the data. This potentially leads to several query interpretations. In this example, SODA creates two interpretations for this query: one where "assets" is an ontology concept and one where "assets" is a data value. Those interpretations are ranked based on an aggregate heuristic function on the lookup scores associated with each lookup result. The initial query interpretations are used to generate additional interpretations. In our previous example, "clients" may be further divided into "individuals" and "organizations" based on the ontology. Those sub-classes could be used to generate two additional, more fine-grained interpretations of the initial query.

BELA (Walter *et al.*, 2012) uses a lexical tree adjoining grammar (Unger *et al.*, 2012) for the semantic tagging step of the input queries. This parsing results in a set of SPARQL query templates, each corresponding to a possible interpretation of the given query. Unlike ATHENA that requires domain-specific ontologies, BELA constructs an inverted index from entities and properties of the cross-domain, publicly available ontology offered by DBpedia (Auer *et al.*, 2007) to fill in the unknown slots in the SPARQL query templates. When lookup results are empty for a slot in a SPARQL query template, BELA retrieves from the ontology of DBpedia the properties of the entities identified in the other slots of the SPARQL template. It uses those properties to find the most similar (via Levenshtein distance) entities from the inverted index, which are associated with the already identified entities for the other slots.

USI Answers (Waltinger *et al.*, 2013) employs Stanford's Core NLP (Marneffe *et al.*, 2006), as well as ClearTK (Bethard *et al.*, 2014) for semantic tagging and NER. In a dictionary- and regex-based lookup step, it produces the candidate entities mentioned in the query, generating different query interpretations. An ontology models the underlying data and is used to determine if there is a relationship between the identified entities mentioned in the query. Similarly to the dedicated annotators that ATHENA uses for detecting nested queries, USI Answers also employs a dedicated relationship annotator for that step, while there are also two more annotators, one dedicated to concepts and one dedicated to temporal mentions.

TR Discover (Song *et al.*, 2015) uses a feature-based context-free grammar for semantic tagging, and also provides query auto-completion. Users start typing a part of the query and they are able to select one of the suggested lexical entries (an entity, or a property) for the current segment of their query. At the same time, TR Discover suggests the next lexical entries that are reachable from the selected query segment, based on the context-free grammar. The ranking of these suggestions is based on the nodes centrality in an RDF graph, in which each node represents a different lexical entry. For example, when users start typing "d", they can select the option "drugs" among the alternatives "drugs using" or "drugs having a primary indication of". Upon selecting "drugs", the users can then select the suggestions "drugs using", or "drugs manufactured by", or "drugs developed by" since those are all properties of the concept "Drug" in the underlying RDF graph. This is a very useful feature that is missing from many other works, as it helps not only the users to understand the query capabilities of the underlying system (by showing suggested queries), but also the NLQ system to have a better control over the user-provided queries.

Dhamdhere *et al.* (2017) propose Analyza, a system which provides a natural language interface integrated into a multi-modal tool for exploring data stored in spread sheets as well as databases. Analyza uses a meta-data store that contains vocabulary used for disambiguation, schema information annotated with column types (measures, dimensions), data formats and ranges, and a knowledge base for mapping entity values like "Germany" to schema elements such as column names like "Country". The system uses a semantic parser based on a context-free grammar that takes the natural language query as input, exploits the metadata store with annotations and extracts the intent, measures, dimensions, filter values such as date ranges, etc. and creates an intermediate structured representation. A simple algorithm is used for identifying the appropriate table in the underlying schema that contains all the columns referenced in the query. A query generator transforms the semantic parse into an actual SQL query or spread sheet formula using a template based generation system. Finally an interpretation generator generates a natural language response to the system's interpretation of the query to be supplied to the user along with the results.

GeoQA (Punjani *et al.*, 2020) is a template-based question answering system over geographic data. It uses Stanford's CoreNLP for part-of-speech tagging and generating a dependency parse tree. For feature types of interest, the Concept Identifier component of GeoQA maps parts of the query to the DBpedia, GADM, and OSM ontologies, using a string matcher, together with a lemmatizer on the entity labels provided by those ontologies, as well as synonyms provided by WordNet. Then, the Instance Identifier component uses the TagMeDisambiguate (Ferragina and Scaiella, 2010) tool for named entity recognition and disambiguation on candidate geographic entities. Some pre-specified properties used in geospatial queries (e.g., height, elevation), associated with specific DBpedia classes (e.g., Mountain), are further employed for better detecting the feature that the user is interested in knowing, once the entity type has been detected from the Concept Identifier. The final query is structured using GeoSPARQL[1], a geospatial extension of SPARQL, from a set of hand-crafted templates. The correct template is identified by comparing the dependency parse tree of the query to the corresponding trees of the query templates.

The rule-based systems described above are strong in their semantic understanding of the input natural language query that is interpreted using different techniques such as indexes, taxonomies and knowledge bases/ontologies. However, they have been shown to be brittle at handling linguistic variations in natural language queries. These limitations can be handled using deep learning technologies such as language models and different neural network architectures that use a data-driven approach to extract relevant context. Next, we describe text-to-SQL systems that use deep learning technologies for generating SQL from natural language.

---

[1]http://www.opengeospatial.org/standards/geosparql

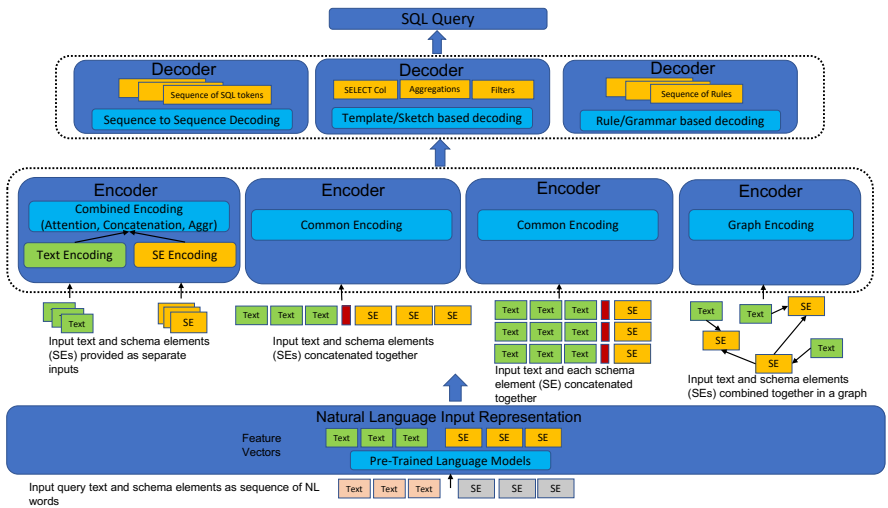

**Figure 3.4:** Text-to-SQL overview.

## 3.2 Text-to-SQL Approaches

Text-to-SQL approaches are an active area of research and development. In this section, we provide a brief overview of these systems and describe a couple of representative systems to illustrate the different techniques and architectures used by these systems for natural language query to SQL translation. For more examples and details, we refer the reader to a recent survey (Katsogiannis-Meimarakis and Koutrika, 2021a), solely focusing on text-to-SQL systems.

The recent success of artificial intelligence and in particular deep learning triggered a new trend of building NLQ systems. The key idea here is to use a supervised learning approach that views the problem of text-to-SQL as a machine translation problem. These systems use a trained ML/DL model that translates a natural language query, a sequence of words, to a structured query such as SQL as an output to be executed against a schema. Figure 3.4 shows an overview of a generic text-to-SQL system architecture with three basic components. The natural language input representation takes the natural language query and schema elements to generate feature vectors. The encoder takes the input feature vectors of the query text and schema elements as input and learns an intermediate representation of this combined input. Finally, this intermediate representation is decoded by the decoder to generate the SQL query. We describe each of these components in detail below.

**Natural Language Input Representation.** The input to a text-to-SQL system is a combination of the natural language query as well as schema elements against which the query is expressed. This input is provided in terms of the natural language query tokens and names of schema elements as a sequence of words, which are converted into feature vectors (or embeddings) using word/sentence embedding techniques described in Section 2.2. These feature vectors are, then, provided as input to the encoder using different structural compositions, which we describe next.

**Encoder.** The input to the encoder, which is expressed as a feature vector of query tokens and schema elements, can be structurally composed in a variety of different ways (Katsogiannis-Meimarakis and Koutrika, 2021a) as shown in Figure 3.4. (1) *Feature vectors of query text and schema elements as separate inputs to the encoder.* Initial text-to-SQL systems such as Seq2SQL (Zhong *et al.*, 2017), SQLNet (Xu *et al.*, 2017) encode the natural language query (a sequence of tokens) and schema elements such as the database column names[2] separately. The encodings are combined using different mechanisms such as attention, concatenation, different types of aggregations, etc. (2) *Feature vectors of query text and schema elements concatenated together.* With the rapid advances of language models, the more recent text-to-SQL systems, such as SQLova (Hwang *et al.*, 2019), SDSSQL (Hui *et al.*, 2021), IRNet (Guo *et al.*, 2019) and ValueNet (Brunner and Stockinger, 2020), use a single encoder for encoding both the natural language query tokens as well as the schema elements represented as a concatenated sequence of words. (3) *Feature vectors of query text and and each individual schema element concatenated together.* This approach used by systems like HydraNet (Lyu *et al.*, 2020) encodes the input query with each schema element like the database column separately allowing for separate predictions for each schema element. (4) *Feature vectors of query text and schema elements combined together as a graph.* Systems like RAT-SQL (Wang *et al.*, 2019) and SMBoP (Rubin and Berant, 2020) use a more complex input encoding wherein the schema elements are represented in the form of a graph preserving relationships between the tables and columns. The natural language query words or tokens are also added to the graph as nodes linked to appropriate schema elements based on entity mappings.

---

[2]Separate encoding allows for column names with multiple words.

**Decoder.** The decoder takes the intermediate representation learnt by the encoder as input and produces the final structured query. As seen in Figure 3.4, there are several different output decoding techniques that are employed by a text-to-SQL system. (1) *Sequence-to-sequence decoding.* Systems like Seq2SQL (Zhong *et al.*, 2017) and SEQ2TREE (Dong and Lapata, 2016) treat the decoding as a sequence-to-sequence translation wherein the model learns to generate the output SQL query as a sequence of tokens. Although viewing the text-to-SQL problem as a sequence-to-sequence machine translation simplifies the architecture, this technique is prone to making syntactical errors, thereby producing incorrect SQL queries that cannot be executed against the database schema. (2) *Template or sketch-based decoding.* Systems like SQLNet (Xu *et al.*, 2017), SQLova (Hwang *et al.*, 2019) and HydraNet (Lyu *et al.*, 2020) break down the problem of generating the SQL query into predicting smaller sub-parts such as predicting the SELECT column values and aggregations, predicates or filters in the WHERE sub-clause, etc. This technique can be considered akin to a template or a slot-filling mechanism which has been shown to work well for simple queries, but difficult to extend to generating more complex SQL queries. (3) *Rule-based decoding.* To enable generation of more complex SQL queries systems like IncSQL (Shi *et al.*, 2018), IRNet (Guo *et al.*, 2019) and RAT-SQL (Wang *et al.*, 2019) generate a sequence of rules instead of a sequence of tokens of a SQL query. These grammar-based rules are applied sequentially to generate the final SQL query.

Although the above mentioned decoding techniques especially template and rule based, have shown to produce SQL quite accurately, there are still instances where these techniques fall short (e.g., an incorrect aggregation on a particular column type or a mismatch while applying a filter condition to a column of a particular type). This could be due to inadequate training data or the complexity of SQL that needs to be generated as such. Execution based decoding (Wang *et al.*, 2018a) avoids making choices that cause such syntactic errors by choosing an incremental mechanism of actually executing partially complete predicted SQL queries to detect faults and exclude those incorrect choices. These choices would however have to be made at the expense of additional prediction time required for execution of partially predicted queries against the database slowing down the inference process significantly. Next, we give a brief overview of some popular text-to-SQL systems that use the above mentioned techniques for input representation, encoding and decoding.

Seq2SQL (Zhong *et al.*, 2017) is one of the earliest text-to-SQL systems which uses GloVe (Pennington *et al.*, 2014) embeddings for input representation. It uses a deep neural network architecture with reinforcement learning to translate natural language to SQL. Seq2SQL uses an LSTM (Hochreiter and Schmidhuber, 1997) encoder-decoder architecture for generating SQL from NL. LSTMs are a type of Recurrent Neural Network (RNN) that use feedback connections i.e. using the output of a cell as input for subsequent steps. Feedback helps build a short term memory (information over a sequence of inputs) that can be persisted over a long time, enabling them to handle sequences of input data and produce sequences of output data. This makes them a good choice of language translation. Further, their use for language translation tasks has gained popularity due to their ability to overcome the problem of vanishing gradients often seen in feedback networks, by using a persistent linear cell surrounded by non-linear layers that feed input and parse output from LSTM cells.

Seq2SQL uses a bidirectional LSTM encoder for encoding the sequence of natural language query, the schema elements (list of all column names) and SQL vocabulary (SELECT, WHERE, COUNT, MIN, MAX, etc.).The input to the encoder are GloVe embeddings corresponding to the input sequence of words. Seq2SQL then applies an augmented pointer network (Vinyals *et al.*, 2015) to generate the SQL query token-by-token by selecting from an input sequence. Seq2SQL takes advantage of the SQL structure and different sub-parts of the deep neural network predict different clauses of the SQL query such as SELECT clause columns and aggregation functions, WHERE clause columns and conditions (shown in Figure 3.5). Specifically, the Seq2SQL network first classifies an aggregation operation for the query, with the addition of a null operation that corresponds to no aggregation. It then points to a column in the input table corresponding to the SELECT column. Finally, the network generates the WHERE clause conditions for the query using a pointer network. The first two components i.e. aggregations and SELECT columns are supervised using cross entropy loss. The generation of the WHERE clause component is done using reinforcement learning to learn a policy to directly optimize the expected correctness of the execution result to address the issue that different orders of conjunctive predicates in the WHERE clause can produce SQL queries with the same result. Finally, Seq2SQL uses a two layer, unidirectional LSTM as a decoder network which uses an attention mechanism to output the final SQL.

*Attention mechanisms* (Vaswani *et al.*, 2017) are widely used in deep neural network encoder-decoder architectures. Similar to cognitive attention, attention mechanisms in neural networks allow them to give more weight to certain parts of the input that is more important than other parts depending on the context. These attention mechanisms manifest as soft weights that vary at runtime and are trained as fully connected neural network layers using gradient decent. Typically a correlation matrix of dot products is used to calculate the attention coefficients that allow the encoders and decoders to give different importance to different parts of input/output based on these soft weights. They are specially relevant for language translation tasks where the encoders for NL input and decoders that produce SQL need to give varying importance to different parts of the input/output sequence based on context.

The Seq2SQL model is trained by using gradient descent to minimize an objective function of the above three components. Namely, the total gradient is the equally weighted sum of the gradients from the cross entropy loss in predicting the SELECT column, from the cross entropy loss in predicting the aggregation operation, and from policy learning for the WHERE clause. Utilizing the structure of SQL allows Seq2SQL to prune the output space of queries, which has been demonstrated to work on simple single-table queries without joins. However, Seq2SQL is limited in terms of the query complexity, because its dependence on the order of input sequence as well as the sequence-to-sequence translation model's ability to only capture rather simple SQL query structures.

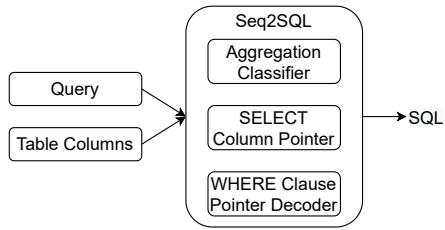

**Figure 3.5:** Seq2SQL architecture (Zhong *et al.*, 2017).

SQLNet (Xu *et al.*, 2017) encodes the natural language query and schema information separately, uses column attention and employs a sketch-based approach to generate SQL akin to a slot-filling task. Each SQL query component is assigned a unique pair of LSTM encoders and column attention is used to ascertain the affinity of words in a natural language query with certain schema elements such as table names. This fundamentally avoids the shortcomings of the sequence-to-sequence approach as the SQL query generation is less dependent on the order of the input and utilizes the more structured sketch-based approach. TypeSQL (Yu *et al.*, 2018a) further improves upon SQLNet by proposing a different training procedure that predicts values for the slots in the SQL sketch in a single model. It further utilizes types extracted from either knowledge graph or table content to help the model better understand entities and numbers in the question.

Other systems that employ a sketch-based decoding approach include HydraNet (Lyu *et al.*, 2020) and SQLova (Hwang *et al.*, 2019). Both HydraNet and SQLova use the pre-trained language model BERT for input encoding. While in HydraNet the input natural language query is encoded separately with each table column, SQLova concatenates all the table columns together and encodes it with the natural language query. In both systems, the BERT output is then decoded for predicting the different parts of the SQL query (slot-filling) such as the aggregations, SELECT columns, filter conditions including columns and operators, etc. HydraNet uses simple linear networks to decode the BERT output and chooses columns for different parts (i.e., SELECT and WHERE) of the SQL query. SQLova uses a more elaborate NL2SQL Layer which further encodes the output of the BERT encoding layer using a pair of LSTM-q (question encoder) and LSTM-h (Table header encoder) along with column attention for each sub-part of the SQL query. In addition, HydraNet also employs Execution-based decoding as proposed by (Wang *et al.*, 2018a) to eliminate SQL errors. Both HydraNet and SQLova only support simple SQL queries with no nested structure and have been shown to achieve good performance on WikiSQL.

Although systems such as HydraNet and SQLova that employ a sketch-based method combined with execution guided decoding perform strongly on the WikiSQL benchmark, the execution-based decoding comes at the cost of significant increase in inference time. Schema Dependency Guided multi-task text-to-SQL (SDSQL) (Hui *et al.*, 2021) proposes a technique to learn the interactions between the natural language query components and schema elements. Similar to SQLova, SDSQL encodes the natural language query and concatenated set of columns using BERT. The output of BERT is then fed to a 2-layer Bi-LSTM to obtain a task-related representation. This representation is then fed into two separate networks, one for SQL prediction and one for schema dependency prediction (i.e., obtaining the dependency between questions and schemas). The schema dependency module uses a Bi-LSTM followed by a multi-layer perceptron (MLP) and a Bi-Affine transformation.

A multi-layer perceptron is a feed-forward network that consists of at least three layers of nodes: an input layer, a hidden layer and an output layer. Each of these layers except for the input have a non-linear activation function such as ReLU. The neurons in the network are trained using back-propagation and the network model approximates a continuous function that can distinguish data that is not linearly separable. Together, the Bi-LSTM, MLP and the Bi-Affine transformation form a bi-affine deep neural network mechanism (Dozat and Manning, 2016), that is widely used in dependency parsing tasks. It decomposes the dependency prediction into the presence or absence of a dependency (edge), and the type of potential edge (label). The SQL prediction module is similar to SQLova. They use an adaptive multi-task loss to optimize both networks and generate dependency training data based on the expected SQL.

SMBOP (Rubin and Berant, 2020) is the most recent work focusing on text-to-SQL semantic parsing. It utilizes the state-of-the-art RAT-SQL encoder (Wang *et al.*, 2019) and introduces a semi-autoregressive bottom-up semantic parser to construct a SQL query at the decoding step. Specifically, at each decoding step $t$, SMBOP uses cross-attention to contextualize the trees (i.e., the SQL abstract syntax tree) with information from the input natural language question. It generates in parallel the top-$K$ program sub-trees of depth $\leq t$. The neural bottom-up parsing also provides learned representations for SQL sub-queries, which are sub-trees computed during the construction procedure, in contrast to top-down parsing, where hidden states represent partial trees without clear semantics.

Other works focus on domain adaptation and transfer learning. Wang *et al.* (2018b) introduce a general purpose transfer-learnable NLQ system with the goal of training a model that could be used against different datasets. The system focuses on the latent semantic structure of natural language queries against relational databases to provide the ability of transfer learning against different datasets. The proposed system separates data instance values and schema elements mentioned in a natural language query using annotations. The annotation process utilizes the database metadata including the schema, database statistics, database values and the knowledge of common natural language expressions used to refer to specific columns of a database. The task of annotating is a two-stage process consisting of mention detection and mention resolution. The first stage detects a set of many possible mentions of column and values, in which some mentions may be inconsistent with others since certain possibilities are mutually exclusive. The second stage finds a maximum subset of these mentions that is consistent, which constitutes the output annotation. The annotated natural language query is translated to an annotated SQL query using a Seq2Seq model. Finally the annotated SQL query is converted to a regular SQL query.

The requirement of large amounts of training data for supervised training of deep neural networks is often a road block while building effective text-to-SQL systems. To alleviate this issue, DBPal (Basik *et al.*, 2018; Weir and Utama, 2019) avoids manually labeling large training data sets by synthetically generating a training set that only requires minimal annotations in the database. DBPal uses the database schema and query templates to describe natural-language/SQL pairs. Specifically, the query templates are used instantiate different possible SQL queries that one might phrase against a given database schema. Namely, the space of all possible SQL queries can be defined using a set of SQL templates. The results show that on a single-table data set DBPal performs better than the semantic parsing approach.

Machine learning-based approaches have shown promising results in terms of robustness to natural language variations. However, these systems still have limited capability of handling complex queries involving multiple tables with aggregations, and nested queries. In addition, they require large amounts of training data, which makes the domain adaption challenging. Next, we describe hybrid approaches that attempt to address some of these limitations.

## 3.3   Hybrid Approaches

Rule-based architectures are more adept at translating more complex natural language queries to complex structured queries. They are also amenable to incorporation of domain knowledge. However, they are brittle to language variations and often require a lot fine tuning to handle such variations. Deep learning based approaches on the other hand are more robust to language variations and are adept at learning directly from data. However, they work well only for simple natural language to structured query translations and often require a large amount of training data. Most of the text-to-SQL systems discussed above do not exploit external knowledge bases or perform any explicit schema linking or entity mapping between the natural language query and the elements of the underlying schema and rather depend on the neural networks to learn these implicitly through training data. This might lead to poor performance when used to query datasets belonging to different domains or datasets containing schema elements (tables and columns) with names not seen while training (out of domain words), resulting in poor domain adaptation.

Hybrid approaches provide the promise of leveraging the strengths of both these architectures to overcome their limitations to build more effective natural language interfaces to data. Next, we describe a few examples of such hybrid approaches.

RAT-SQL (Wang *et al.*, 2019) is one of the text-to-SQL systems that employs an encoder that takes a more complex input in the form of a graph (depicted in Figure 3.6). The graph is created using explicit schema linking techniques that leverage information from external sources making the system hybrid. It creates a heterogeneous question contextualized schema graph wherein the nodes represent both schema elements as well as entities or tokens from the natural language query. This is done in a two step process. First, the database schema is represented as a directed graph wherein nodes represent columns and tables each labelled with the their corresponding schema names. The edges between nodes in the graph that represent schema elements are inferred using the schema relations such as the primary-key foreign key relations. This is a one time process for each database schema.

Second, for each NL query, the query tokens are introduced in the graph as nodes and the edges between the nodes representing the natural language query tokens and the nodes representing the schema elements are inferred using explicit schema linking including schema name-based linking (using exact or partial n-gram match) as well as value based linking (using indices or textual search). Specifically, name-based linking refers to exact or partial occurrences of the column/table names in the question, since textual matches are the most explicit evidence of question-schema alignment. Value-based linking, on the other hand, becomes useful when the question mentions any values that occur in the database and consequently participate in the desired SQL. RAT-SQL captures a column-value relation between any word in a natural language query and column name if the word occurs as a value or a full word within a value of the column name.

The initial input representation for input encoding is provided using GloVe, or LSTM, or BERT. Every RAT layer uses self-attention between all elements of the input graph to compute new contextual representations of question words and schema members. This self-attention is biased towards some pre-defined relations using the edge vectors in each layer. The set of used relation types are predefined using the name-based linking and value-based linking described above. Occurrences of these relations between the question and the schema constitute the edges.

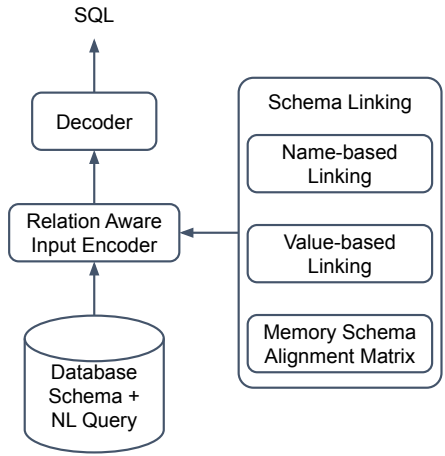

**Figure 3.6:** RAT-SQL framework (Wang *et al.*, 2019).

RAT-SQL employs a modified transformer architecture (Vaswani *et al.*, 2017) that uses relation-aware self attention for the network to learn the biases towards the different types of relationships in the input encoding. The transformer architecture consists of an encoder and decoder. However, instead of using LSTMs to capture the sequences, the encoder and decoder mainly consists of several stackable units of multi-head attention and feed-forward layers. Both the input (NL query token embeddings) and output (target SQL query token embeddings) are embedded into a latent space. Since there are no LSTMs to capture the sequence, an important addition to this architecture are *Positional Encodings*. These positional encodings allow the specification of the relative position of every token in the input and target sequence which depends on the order of its occurrence in the sequence. These positions are added to the token embeddings of each word in the NL query and target SQL query. Each of the multi-head attention layers in the encoder and decoders compute an attention matrix by multiplying the vector representation of each word (Q) in the input sequence with the vector representations of all the words in the sequence (K, also called Keys). Therefore, the weights in the attention matrix capture how each word of the sequence is influenced by all the other words in the sequence. A softmax function is applied to scale the weights between the values of 0 and 1.

Finally, the vector representations all the words in the sequence (V, also called Values (Same as Q)) are multiplied by these weights. The attention mechanism is repeated multiple times to allow the system to learn from different representations of Q, K and V. These linear representations are done by multiplying Q, K and V by weight matrices W that are learned during the training. The Feed Forward Layers after the attention layers allow for a linear transformation of each element from the given sequence. The output of the encoder is fed as input into one of the multi-head attention layers of the decoder so that the encoder input-sequence is taken into account together with the decoder input-sequence. Note that in this case the input V to the multi-head attention layer in the decoder is the output of the previous attention head in the decoder. The final output of the decoder is a softmax layer which outputs probabilities. Given an input encoder sequence and a particular decoder output sequence shifted by one, the transformer model learns to predict the next word in the sequence.

In RAT-SQL, the relation-aware transformer is directly followed by a grammar-based LSTM decoder which produces a sequence of rules or decoder actions used to construct an abstract syntax tree similar to IRNet. The final SQL is inferred from the generated abstract syntax tree.

Approaches such as Ben Abacha and Zweigenbaum (2015), Bast and Haussmann (2015), and Bergamaschi *et al.* (2016) combine rule- and learning-based query understanding in a multi-step strategy making them hybrid approaches. For example, QUEST (Bergamaschi *et al.*, 2016) first chooses the entities that are relevant to the keywords in the query based on Hidden Markov Models (HMM), trained on a data set of previous searches, validated by the user. The relationships between the entities extracted from the query are then computed based on heuristic rules that consider the relationships of those entities in the database. The candidate interpretations are ranked based on the aggregate confidence scores returned by the HMM. However, these systems are still not capable of covering a full spectrum of the complexity of generated queries. Hence, more research is needed on this hybrid approach that attempts to leverage the best from both worlds.

Other approaches such as IRNet (Guo *et al.*, 2019), address the issue of better domain adaptation with an explicit schema linking mechanism between the tokens of natural language question and the elements of the database schema (tables, columns), thus making them hybrid. All n-grams of length 1-6 in the natural language query are tested to match against the database schema resulting in some complete and partial matches. For data instances or values the system uses ConceptNet (Net, 2021) to determine the corresponding schema element to link to. IRNet uses BERT encodings wherein the natural language spans are appended with their linked schema tokens. IRNet uses a grammar based decoder model to generate an intermediate representation in terms of a SemQL (Semarchy, 2021) abstract syntax tree. The systems then infers the SQL query from this intermediate representation using domain knowledge obtained through explicit schema linking.

ValueNet (Brunner and Stockinger, 2020) further improves upon IRNet by improving the condition value prediction for the filter clauses in a SQL query. Often times the values mentioned in the natural language query do not match the values stored in the database. This could be because of the use of abbreviations such as IBM for "International Business Machines" or use of synonyms of values stored in the database. ValueNet uses NER along with some heuristics for value extraction from the natural language query. It then generates candidate values from the database that are similar to the extract entities using indices and string pattern matching. The input encoding concatenates the natural language query with the schema elements (Table and column names) as well as the identified candidate values. ValueNet uses a similar decoder architecture as IRNet with an improved SemQL 2.0 grammar (Lee and Baik, 1999) to generate the abstract syntax tree and infer SQL queries.

DBTagger (Usta *et al.*, 2021) provides a deep learning architecture based on bi-directional Gated Recurrent Units (GRUs) (Cho *et al.*,

2014) for the task of entity tagging (Figure 1.1) in NLQ pipelines. It extends the supervised learning DL architecture of sequence tagging, by utilizing multi-task learning and cross-skip connections. They learn POS (part-of-speech), type and schema tags from annotated query logs, and make the observation that the schema tags of keywords are highly correlated with POS tags. Their techniques can be plugged into any rule-based NLQ pipeline to identify the mapping of the keywords in the user query to the database elements.

**Summary**

In this section, we present two commonly used paradigms for natural language interfaces to data, rule-based systems and text-to-SQL systems. Rule-based approaches rely on a semantic data model, such as an ontology, knowledge graph, or a semantic index to identify entities and their relationships in the user's natural language query, and use rule-based interpretation techniques to generate the final SQL or SPARQL query. Text-to-SQL systems, on the other hand, encode user input into a feature embedding and train deep learning models to generate the SQL query in a holistic way. These systems leverage the recent advances in language models and NLP to convert input texts into SQL. While rule-based approaches provide easier domain adaptation, text-to-SQL systems are more robust to the paraphrasing of the input query. There are also a few emerging hybrid approaches that leverage the best of both worlds to overcome their limitations to build more effective natural language interfaces to data. Hybrid approaches use a combination of DL models and rule-based techniques for different problems, such as entity tagging or structured query generation, in the NLQ pipeline. We believe more research is needed for all these techniques to provide better accuracy and reliability, as we discuss in Section 6.

# 4

## Conversational Data Analysis and Exploration

Conversational systems are an extension of natural language interfaces to support a two-way conversation (a dialogue) between the user and the system, using the principles of human-to-human conversation (Gao *et al.*, 2018). Conversational interfaces to data are rapidly gaining popularity because of their unique ability to enable exploration of data and derivation of insights in small incremental steps as the conversation with the data progresses. Conversational systems can understand, respond and clarify ambiguity through interactions with the user in natural language, while persisting the context of the conversation across multiple turns.

The technologies exploited by NLQ systems for natural language understanding, generation of structured queries to retrieve data and natural language responses, as mentioned in Section 3, are also applicable for building conversational interfaces to data. However, this requires extending their ability to represent the two-way structure or multi-turn dynamics of a dialogue. Our focus in this section is to describe systems and technologies that enable this extension to conversational systems by exploiting models that capture both the multi-turn dynamics as well as the structural context in a dialogue. We limit our discussions specifically to technologies and systems pertinent for building conversational interfaces for data exploration. These systems could be considered to fall under the category of task-based conversational systems where the task is data exploration/analysis. We cover both open-domain and domain-specific systems (as classified earlier in Figure 2.3). For a more comprehensive overview of the basic components of conversational systems and the challenges involved in the application of deep learning techniques for the development of state-of-the-art dialogue systems, we refer the reader to (Chen *et al.*, 2018).

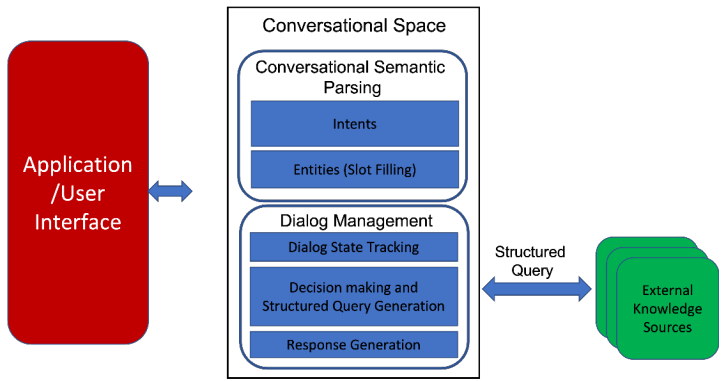

**Figure 4.1:** Conversational system: technology stack.

Figure 4.1 shows an overview of the technology stack employed for building conversational interfaces to data. This includes conversational semantic parsing and dialogue management. Conversational semantic parsing provides the necessary capability of language understanding. Similar to the language understanding techniques mentioned in Section 3, conversational semantic parsing entails parsing of natural language queries/utterances across multiple turns of conversation for detecting intents and entities. Intent identification is usually framed as a user utterance classification problem (Chen *et al.*, 2018), wherein each user query can be classified to a particular intent with a certain probability. Identifying entities is more traditionally referred to as *semantic slot filling* in conversational systems, i.e., identifying (tagging) words/tokens in the natural language text with their appropriate semantic entity types (see Section 2.3, Semantic tagging). Together, conversational semantic parsing provides a more structured semantic representation of the user input enabling natural language understanding.

Dialogue management includes (1) state management to track the current state of data exploration, given the history of prior interactions between the user and the system, (2) decision making and structured query generation for choosing an appropriate external knowledge source and retrieving data for a given user query, and (3) NL response generation to describe the results obtained and enable a two-way conversation with the user. Next, we describe each of these technologies and their state-of-the-art in further detail. The section also introduces a use case for conversational interfaces to data for business intelligence systems and applications.

## 4.1   Conversational Semantic Parsing

Semantic parsing of natural language utterances enables the extraction of a structured semantic representation of the user utterance that enables inferring of intent and entities. We discuss below some of the challenges associated with each, as well as those that arise while extending capabilities of semantic parsing to a two way dialog.

Intent identification is modelled as a classification task and requires labelled training data for training classification models such as Support Vector Machines (SVMs) or more recently Deep Neural Networks. Some of the challenges involved in training classification models for intent identification involve identifying the number of intents(labels) the model can support with high accuracy. This requires the estimation of the actual distribution of queries in the expected workload and is often determined empirically. Dealing with class imbalance during training is also a frequently encountered problem while training classification models on real data, as real workloads are often skewed. This may lead to classification bias towards the majority class, skewing the classification boundary and over fitting. Techniques such as blocking which under sample the majority class to reduce imbalance, customizing the loss function to penalize mis-classification of minority classes, data augmentation (both real and synthetic) to reduce class imbalance are some of the techniques that are used to overcome these limitations.

Identification of entities or slot filling also faces several challenges that have been addressed using several different approaches. These include rule-based approaches, such as those discussed in Section 3.1, that employ a context free grammar to parse a user utterance and identify the non-terminals as the associated tags. Although these approaches have the advantage that they are very precise, they involve a lot of manual effort in defining the set of rules required. Machine learning approaches address this problem and require feature engineering to represent words in text and use a data corpus to train probabilistic models that provide the probability of tags associated with a word. Deep learning approaches further improve on these ML techniques as they do not require manual feature engineering. Encoded words (word embeddings) are fed to a deep neural network that extracts the required features to predict the tags associated with the words in the user's utterance.

Additional challenges for conversational systems arise from extending the capabilities of traditional semantic parsing technologies that are designed for understanding the structure of single user utterance or one-shot queries. These challenges include dealing with co-reference resolution (i.e., references to the same entity) and context from prior queries in the session that are traditionally handled by the dialogue system. Conversational semantic parsing addresses this limitation of semantic parsing of single user utterances, and extends the concept of semantic parsing of a natural language query to a sequence of natural language queries issued by the user across multiple turns of conversation. This allows for co-reference resolution and context carryover from prior queries for appropriate semantic slot filling, enabling a comprehensive understanding of user queries in a given data exploration session. Next we describe a few state-of-the-art deep learning approaches for conversational semantic parsing that address some of the challenges associated with context representation across multiple turns of user utterances.

**Seq2Seq Approach.** Aghajanyan *et al.* (2020) propose a Seq2Seq approach for a session-based parsing that attempts to extend the existing semantic parsers for one-shot query answering to conversational semantic parsing for multi-turn queries. They enhance the traditional Seq2Seq architecture based on the Pointer-Generator architecture (See *et al.*, 2017) to build a semantic parsing model that removes the tight coupling between the sequence of words in the original user utterance and its semantic representation, focusing more on the semantic entities (slots) and removing text snippets occurring in a non-leaf spot in a regular compositional semantic parse. This allows words to be grouped logically into the same slot even though they may be spread apart in the original sentence. For example, in the utterance *"For financial year 2021, show me the net revenue in the first Quarter for Microsoft Inc."*, *financial year 2021* and *first quarter* belong to the same date-time slot. Their proposed model provides further enhancements to the decoupled semantic parse model with the use of a bidirectional LSTM, or a transformer to encode a sequence of user queries. The decoder produces a semantic parse tree structure that captures relevant information across the user utterances in a session. This allows for co-references which are explicit references to slots in a previous utterance, and implicit references based on contextual information (slot carry-over), enabling a better understanding of information across user queries in a session especially relevant for data exploration.

**Pre-trained Language Model-based Approach.** Another approach for context representation across multi-turn queries in conversational semantic parsing is based on pre-trained language models. Current language models pre-trained on free-form text corpora are limited in their ability to represent the two-way structure or multi-turn dynamics of a conversational system for data exploration. SCoRe (Structured and Sequential Context Representation) (Yu *et al.*, 2021) addresses the limitations of existing language models (Section 2.2) for conversational semantic parsing by introducing a second phase of pre-training that captures the two-way conversational patterns as well as the dialogue context across multiple turns. The second phase of pre-training consists of training a task-oriented language model contextualized by conversational flow and the underlying schema/ontology. In addition to masked language modeling for contextual representation learning of natural language utterances, SCoRe utilizes multiple training objectives: (1) *column contextual semantics:* that enable linking of user utterances to schema elements of an underlying ontology (slot filling) and (2) *turn contextual switch:* that captures the relationship between different ut-terances across multiple turns of conversation. Together, this enables the model to learn an accurate representation of the conversational flow across multiple turns and its mapping to the underlying schema.

## 4.2 Dialogue Management

The dialogue defines the space of interaction patterns supported by a conversational system (Section 2.5). The dialogue subsystem provides a natural language response to a user conditioned on the identified intents, extracted entities (slot filling) in the user's input using conversational semantic parsing described above. It also takes into account the current context of the conversation persisted across multiple turns of the conversation (i.e. state of data exploration) as well as the results obtained from external knowledge sources through the execution of a structured query. This carrying over of context from prior user queries is what essentially differentiates dialogue from one-shot query answering.

We describe the different components of dialog management, including (1) dialog state tracking, (2) structured query generation, and (3) natural language response generation (Figure 4.1) in further detail below. Note that although we describe each of these tasks separately and provide references and expositions of example systems that implement these tasks using different techniques, these individual tasks are closely intertwined with each other. With recent advances in machine learning, particularly in deep neural networks, there is an increasing trend in using these techniques to learn dialogue management models as a whole that can accomplish all of these tasks.

### 4.2.1 Dialogue State Tracking

Dialogue state tracking enables estimating the current state of the dialogue (also referred in literature as the *belief state*) given the history of prior conversation between the user and the system. This entails keeping track of the current *state of data exploration* given the prior set of queries issued by the user in a data exploration session[1]. The system uses this internal representation of the current state of data exploration obtained via dialogue state tracking to decide on the next action such as issuing a structured query against an external data source to obtain results and providing an appropriate natural language response to the user.

---

[1]We define a data exploration session as the set of queries issued by a user for a given analysis task.

Several different approaches have been proposed for building the dialogue structure and maintaining state across turns for a conversational interface: (1) *Rule-based systems* (McTear, 2002; Mallios and Bourbakis, 2016) are used in finite-state dialogue management systems which are simple to construct for tasks that are straightforward and well-structured, but have the disadvantage of restricting user input to predetermined words and phrases; (2) *Frame-based systems* (Fitzpatrick *et al.*, 2017; Beveridge and Fox, 2006; Giorgino *et al.*, 2005) address some of the limitations of finite state dialogue management by enabling a more flexible dialogue. Frame-based systems enable the user to provide more information than required by the system's question, while the conversation system keeps track of what information is required and asks questions accordingly; and (3) *Agent-based systems* (Bing-Hwang Juang and Furui, 2000; Young *et al.*, 2013; Radziwill and Benton, 2017; Miner *et al.*, 2016) that are able to manage complex dialogues, where the user can initiate and lead the conversation.

Agent-based methods for dialogue management are typically data driven statistical models trained on corpora of real human-computer dialogues, offering robust contextual natural language understanding across multiple turns of a conversation, as well as better scalability and greater scope for adaptation. Hence, these are ideally suited for dialogue state tracking for iterative data exploration driven by the user, allowing estimation of the current state of data exploration based on what the user has already explored. Deep learning based techniques for dialogue state tracking fall under the category of these agent based systems. Some of the major challenges associated with state tracking based on trained models include dealing with unseen mentions i.e., utterances not seen by the state tracking model during training, domain adaptation to account for lexical variations, and building the appropriate semantic context to represent the conversation state for more complex dialogue domains. Next, we describe a few deep learning based models and systems for state tracking that address these challenges.

**Dealing with Unseen Mentions.** Traditional agent-based machine learning models that employ generative and discriminative approaches for dialogue state tracking are unable to deal with unseen mentions (Williams and Young, 2007) due to their reliance on a fixed underlying schema/ontology. On the other hand, reading comprehension-based techniques require the determination of answer spans within a given piece of text. These models are developed in a manner that a fixed ontology for an answer is not imperative. Hence, they do not require a fixed vocabulary for response generation (Reddy *et al.*, 2019).

Taking advantage of these developments in reading-comprehension-based techniques, Gao *et al.* (2019) formulate the dialogue state tracking problem as a reading comprehension task and propose an attention-based neural network model that derives the appropriate slot values (mappings to schema or ontology) from a conversation. More specifically it employs three sub-models to accomplish this task: (1) The slot carry over model, that predicts the population of a slot value from the previous conversational turn (context). (2) The slot type model, that predicts whether the slot value is a named entity found within the dialogue, and (3) the slot span model, that predicts the span of the slot value identified in the dialogue.

Rastogi *et al.* (2017) propose a scalable architecture for handling entities not seen during training. Instead of representing the dialogue state as a distribution over the value set for each slot, pre-specified in a fixed ontology, they choose a deep learning based approach that represents slots with large or unbounded sets of possible values. The proposed dialogue state tracking model estimates a set of slot value candidates based on the local conversation context, and can incorporate external knowledge sources. Operating on these candidates instead of a fixed ontology, allows dealing with unseen mentions and scales dialogue state tracking to much larger and richer datasets.

**Domain Adaptation for Lexical Variations.** To address the problem of linguistic variability across different domains, deep learning approaches for dialogue state tracking focus on a tight coupling between language understanding and state tracking. RNN (Sherstinsky, 2020) based models such as (Henderson *et al.*, 2013) and (Henderson *et al.*, 2014) provide a tight integration between the language understanding provided by conversational semantic parsing and state tracking across multiple turns of conversation. They have been shown to improve upon more traditional Bayesian approaches (Thomson and Young, 2010). These models can handle language variability through the provision of dictionaries that capture domain specific lexical and morphological variations. Although these dictionaries handle domain adaptation well, these approaches do not scale well to more complex and larger dialogue domains.

Mrksic *et al.* (2016) address the problem of adaptation to larger and more complex dialogue domains by using neural dialogue models to avoid the need for manual dictionaries to match the lexical variations. Their proposed model uses pre-trained language models to generate word embeddings that are used for identifying the entity types (schema mapping) to populate and track the appropriate dialogue state. The language models do away with the requirement of exact matching of words and their lexical variations while catering to the different linguistic variations in user utterances while exploring data from different domains.

**Deep Domain Specialization.** Although the above mentioned systems that provide domain adaptation for lexical variations, enable dialogue state tracking across larger and multiple domains, they are mostly trained over open domain datasets and are generally domain agnostic. Models trained over such domain agnostic data lack the deep domain specialization required in several real world use cases. These include conversational systems for many domain-specific data sets and knowledge bases, that are carefully curated from various data sources, and serve as a valuable reference for professionals in different domains. Building conversational interfaces for such datasets requires a system to have deep domain specialization for natural language understanding to identify intents and enable mapping of domain-specific user utterances to appropriate entities in the underlying knowledge base or dataset (Domain specific slot filling).

Quamar *et al.* (2020a) propose an ontology-based conversational system for domain-specific knowledge bases. Their proposed system exploits the domain knowledge captured in an ontology representing the knowledge base schema in terms of relevant entities as ontology concepts, their linguistic variations as domain specific synonyms and relationships to other entities (hierarchical, functional, unions, etc.). They use a tightly coupled semantic parsing and dialogue state tracking architecture to identify user intents as patterns (sub-graphs) over the ontology, and domain-specific entities in user utterances as ontology concepts. More specifically, they use an Ontology Page Rank algorithm to identify key concepts that are often the main focus of user queries and dependent concepts that can be mostly seen as complex attributes of the key concepts. They define several query patterns over these identified key and dependent concepts in the ontology as intents. They develop a framework that uses this information in the ontology to bootstrap the intents, entities as artifacts of the conversational interface and allows for incorporating feedback from domain experts to further refine the identified intents/patterns and entities. Since intent identification from user utterances is modelled as a classification task, the proposed framework also automatically generates training samples for training a neural network-based intent classifier.

### 4.2.2 Structured Query Generation

Structured query generation is one of the actions that the dialogue system needs to take to reach a particular dialogue state. This happens as a result of state tracking in response to a user query, or a set of queries to enable retrieval of results from an external knowledge source. Several different techniques could be employed to generate structured queries like SQL. Simple template-based mechanisms map each identified intent to a structured query template and use appropriate entities tracked in the dialogue state to fill in the template to generate an executable query (Quamar *et al.*, 2020a). This is suitable when the number of intents are small and can lend themselves to a manual/semi-automated mapping of intents to structured query templates.

Many of the deep learning text-to-SQL techniques (discussed in Section 3.2) are also applicable in the conversational settings for generating structured queries like SQL in response to natural language utterances issued by the user and tracked by the dialogue system. However, the conversational setting has the added benefit of allowing the system to incorporate user feedback to refine the structured query if required.

DialSQL (Gur *et al.*, 2018) is a dialogue-based structured query generation framework that leverages a human-in-the-loop to boost the performance of existing structured query generation algorithms via user interaction. They introduce a goal-oriented dialogue model that interacts with users to extract and correct potential errors in the generated structured queries. DialSQL is based on a hierarchical encoder-decoder architecture and uses attention and pointer mechanisms. The model encodes each turn of conversation and uses a RNN network across turns in the dialogue history. The output of the RNN network is then used to predict the error category in terms of the SQL clause such as selection, projection, or aggregation error. A second RNN, whose output is conditioned on the identified error category, is used to predict the error span. Lastly, these candidate choices are decoded from the error category and span representations as potential errors in the generated SQL query. Users are then asked for validation via simple multi-choice questions and user feedback is used to revise the query accordingly.

Zhang *et al.* (2019) propose SQL query generation by editing the query in the previous turn. The previous query is first encoded as a sequence of tokens, and the decoder computes a switch to change it at the token level. This sequence editing mechanism models token-level changes and is thus robust to error propagation. Furthermore, to capture the user utterance and the complex database schemas in different domains, an utterance-table encoder is used based on BERT (Devlin *et al.*, 2019) to jointly encode the user utterance and column headers with co-attention, and a table-aware decoder is adopted to perform SQL generation with attentions over both the user utterance and column headers.

Lyons *et al.* (2016) propose Echo Query, that allows querying databases using a hands free voice based natural language interface for dialogue interaction. It is built using the Amazon Alexa Voice Service (Amazon, 2018). It uses user feedback for query clarification and iteratively builds and personalizes the system vocabulary through multiple user interactions. The system focuses on natural language to SQL translation for simple SELECT, PROJECT, JOIN queries with filters and GROUP BYs.

Having described a number of different techniques for generating structured queries, it is important to note that evaluating the quality of generated SQL or structured queries in terms of correctness, consistency and confidence is still an open challenge, with on going work on developing relevant benchmarks for the same. We discuss some of these issues in Section 5 and Section 6.

### 4.2.3 Natural Language Response Generation

The third and last task in dialogue management is natural language response generation. Natural language response generation is conditioned on the input user utterance (identified intent, entities using conversational semantic parsing), the conversational context in terms of information persisted from prior queries (state tracking), the resulting action in terms of a structured query executed against an external data and the data retrieved as a result of such a structured query execution. Natural language response generation is thus tightly coupled and dependent on the other tasks of dialogue management. Next, we describe some of the techniques for generating natural language responses in conversational systems.

**NL Response Generation using Seq2Seq Models.** Deep learning systems use the contextual information captured using dialogue state tracking as input to the encoder, while the decoder is used to generate the next dialogue response in natural language. Wen *et al.* (2015) propose an NLG technique for generating the next dialogue response using a semantically conditioned LSTM that learns from input data by jointly optimizing sentence planning and surface realization, i.e. the task of generating the linear form of a text following a given grammar. Typically, surface realization models consist of a cascade of complex neural network-based sub-modules, each responsible for a specific sub-task. The proposed generation model by Wen *et al.* (2015), is a RNN based encoder-decoder architecture where the input tokens are 1-hot encoded and the output sequence of tokens are lexicalised to form the required NL utterance. The system uses Semantic Controlled LSTM cells where the upper part is a traditional LSTM cell responsible for surface realisation, while the lower part is a sentence planning cell based on a sigmoid control gate and a dialogue action (DA).

Seq2Seq models with attention have also been used for next dialogue generation in conversational systems (Dusek and Jurcicek, 2016). The input to their generator are actions such as *inform* or *request*, along with one or more attributes (slots), and their values, providing context to the actions generated by the dialog state tracking model. The generator produces deep syntax trees that provide the syntactic shape of the output corresponding to the sentence planning NLG stage. These trees are then linearized to strings using a surface realizer. The actions, deep syntax trees, and corresponding sentences are represented as a sequences of tokens as input to an RNN based Seq2Seq generator model with attention. The model uses an encoder-decoder architecture with beam search to generate the natural language output. Finally a reranker ensures that the output strings correspond semantically to the input action. For e.g. for an *inform* action about a particular entity X could lead to a NL sentence generation such as *X is a restaurant.*

**NL Response Generation using Language Models.** As described in Section 2.2, generative pre-training based language models like GPT-2 trained on very large datasets have shown remarkable success in capturing long term dependencies in textual data and generate text with consistent content and style. This makes them an attractive choice for neural response generation for conversational systems. Inspired by the success of GPT-2, DialoGPT (Zhang *et al.*, 2020a) exploits GPT-2 to overcome the challenges of conversational neural response generation i.e. generating natural language text that is relevant to a given prompt. DialoGPT is formulated as an autoregressive language model (a language model which predicts future values from past values) and employs a multi-layer transformer architecture which enables long-term dependency information to be better preserved over time. It is trained on dialogue sessions from Reddit. DialoGPT models the multi-turn conversational sessions as a long concatenated text of dialogue turns and frames the generation task as language modeling. The conditional probability of generating a target response given the sequence of dialogue turns is modelled as a product of conditional probabilities of generating individual target tokens given the sequence of dialogue turns.

TOD-BERT (Wu *et al.*, 2020) uses pre-trained language models, like (BERT) trained specifically on multi-turn task oriented dialogue datasets and exploits a contrastive objective function. A Contrastive objective function allows models to learn semantic representations of input data such that two similar items are placed closer together in the latent space while two dissimilar items would be placed further apart. TOD-BERT uses the contrastive objective function for training the model to simulate the response selection task i.e. representations of an input and its relevant response are closer together than irrelevant responses. It uses a masked language model and a deep bidirectional transformer encoder wherein the dialogue behavior is modelled using additional tokens that distinguish the user and system utterances. Using the above described architecture and additional tokens, TOD-BERT has been shown to do well for the task of appropriate response selection based on the dialogue state.

ConveRT (Henderson *et al.*, 2019a), similar to DialoGPT, is also trained on conversational sessions (Reddit data) for the general purpose response selection task i.e., choosing the most appropriate response given the dialogue history. This is followed by model fine-tuning using additional neural network layers for task-specific response selection. ConveRT employs a dual-encoder pre-training architecture that leverage sub-word representations, transformer style-blocks and quantization. The dual-encoder architecture consists to two parallel encoders with shared parameters, one for the input sub-words and one for the response sub-words. Each encoder transforms the corresponding sub-words into embeddings using a pre-trained language model which are then passed through a series of transformations based on a standard transformer architecture wherein positional encodings are added to the sub-word embedding inputs before going through the self-attention blocks. The final layer in both encoders is linear which maps the input text and response text to L2-normalized vector representations. Cosine similarity is then computed between input and response sub-words with the aim of keeping a pair of input and its relevant response vectors closer together in latent space and irrelevant responses further apart from the input vector. Such dual encoders have been shown to perform extremely well for response selection/generation tasks in conversational systems (Henderson *et al.*, 2019b) for data exploration.

## 4.3   Conversational Business Intelligence

In this section, we describe conversational Business Intelligence (BI) systems which play a critical role in decision making in the enterprise. These interfaces democratize access to valuable business insights from the underlying data for a wide range of personas including non-technical business owners, executives and data scientists, allowing them to explore data, investigate key performance indicators using natural language queries without relying on external technical expertise. We begin with describing some of the challenges in building conversational BI systems. Next we explore the support for natural language interfaces in existing commercial BI tools and platforms and then describe some conversational interfaces specifically built for supporting BI applications.

Building effective conversational systems for business intelligence applications requires: (1) A rich semantic data model that enables natural language understanding in terms of the entities, relationships and associated semantics that are relevant for typical BI workloads such as measures (quantifiable attributes), dimensions (categorical attributes), their hierarchies, relevant aggregations as well as relationships between the measures and dimensions. Such information is typically available in an OLAP (Online Analytical Processing) cube definition (Chaudhuri and Dayal, 1997) over the underlying data. (2) Building the capabilities of the conversational system through suitable models and training data to recognize, interpret and respond to typical BI query patterns/operations as well as the ability to recognize and support OLAP sessions, the typical unit of data analysis for BI use cases. (3) Integration with the underlying BI platforms through generation of appropriate structured queries or API calls to provide a response to the user utterance, often, in terms of a visualization.

Existing business intelligence tools, such as Microsoft's (*Power BI Platform* 2021), Tableau's (*Ask Data | Tableau Software* 2021), (*MicroStrategy* 2021), Amazon's QuickSight (Amazon, 2021), Google's Looker (Google, 2021a) and IBM's (*Cognos Assistant* 2021), support some form of natural language interfaces to their respective BI systems. Most of these systems typically support a fixed set of query patterns. Some of these systems also help the users to complete their current query

with a simple list of query suggestions. These systems rely on the user to formulate a one-shot query by selecting from a large number of options and parameters. The user often gets to an appropriate visualization with limited system support from a two-way dialogue.

Kuchmann-Beauger *et al.* (2013) propose QUASL, a framework for question answering over a multi-dimensional model targeting the BI use cases for the enterprise. The framework requires an explicit mapping (Schema Linking) between the entities expected in the natural language queries and the entities in the underlying data relevant for the BI queries. The framework, however, lacks the ability to clarify ambiguity in natural language and also does not support the use of context across multiple turns of conversation, and hence is limited in its ability to recognize analytical tasks typically carried out in OLAP sessions.

Francia *et al.* (2020) propose an approach that utilizes a Knowledge Base that stores the OLAP cube definition meta-data and values. More specifically, these include measures, dimensions, their hierarchies, aggregation operators etc. The information in the knowledge base is referred to by the system for entity identification (slot filling), query intent interpretation as well as structured query generation. Additional synonyms are added using open data ontologies for better understanding of the natural language queries and accommodating linguistic variability. The conversational interface is also used to clarify any ambiguities in the user query. The system however lacks in modeling the BI query patterns observed in typical workloads and relies on intention keywords such as SELECT, GROUPBY, FILTER, etc., in user utterances. Further, there is no specific dialogue support for handling commonly observed BI operations such as pivot, drill-down across turns of conversation.

Quamar *et al.* (2020b) propose an ontology-driven conversational system for business intelligence (BI) applications. Figure 4.2 shows the workflow for building the conversational system. A BI ontology is used to provide deep domain specialization, reasoning capabilities as well as rich semantics over the domain schema in terms of measures, dimensions, their hierarchies and relationships as defined in the OLAP cube definition over the raw data.

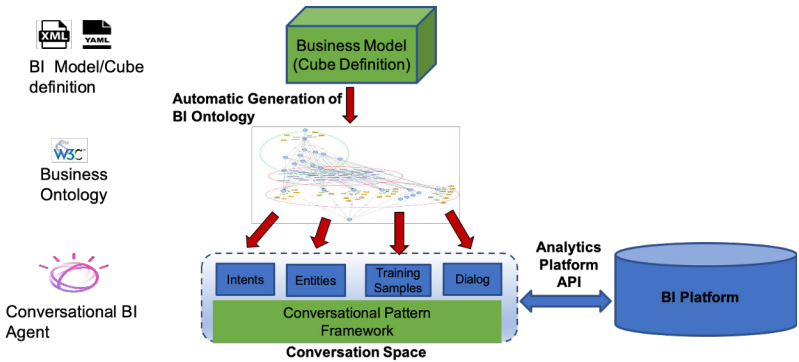

**Figure 4.2:** An ontology driven approach for conversational BI systems (Quamar *et al.*, 2020b).

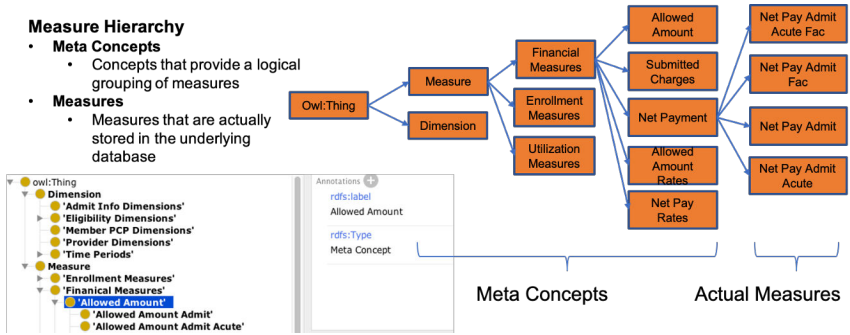

**Figure 4.3:** Captured measure hierarchy(Quamar *et al.*, 2020b).

In addition, the ontology is further enriched by providing a logical grouping of measures and dimensions under higher-level concepts (meta-concepts) that provide richer semantics for natural language understanding (Figure 4.3). This enrichment is done either manually with the help of domain experts or learnt from the underlying data distributions using ML/deep learning techniques. The BI ontology provides domain-specific natural language understanding capabilities for the conversational system for identifying entities (slot filling) in terms of measures, dimensions, their hierarchies and intent in terms of BI workload patterns over these entities and relationships between them. Figure 4.4 shows an example BI workload pattern that compares two measures along a particular dimension. Each such pattern is identified as an intent. The framework maps these BI workload patterns onto the BI ontology and automatically generates training samples to train deep neural networks for intent classification.

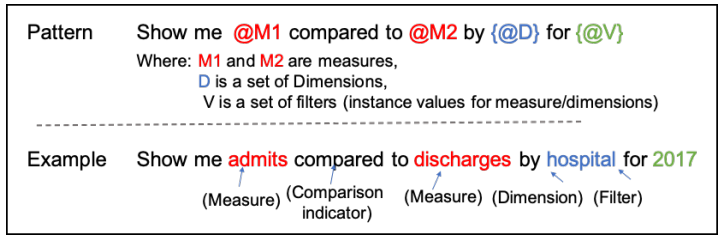

**Figure 4.4:** BI comparison pattern (Quamar *et al.*, 2020b).

The system also provides a tight coupling between conversational semantic parsing for natural language understanding and domain specific dialogue state tracking. The dialogue state tracking supports the desired interaction for the application in terms of BI operations conditioned on identified intent and entities. This requires the state tracking model to keep track of the state of all required entities for a given intent which is modelled as a BI access pattern. If the current data exploration context contains all the required entities (such as measures, dimensions, filters, etc.) for the identified BI pattern in the user's utterance, the state is marked complete and further action can be initiated in terms of forming a structured query and retrieving results. Further, the dialogue exploits context over multiple turns of interactions with the user to support OLAP sessions and the typical operations carried out within such as drill-down, roll-up, pivot, etc. Interaction with the external data source (or BI platform) is done through generated structured queries using a template based mechanism (Section 4.2.2) that exploits BI patterns to retrieve data in response to user/application queries. Each BI pattern is mapped to a structured query template. The template is populated with the identified entities (measures, dimensions, filter values, aggregation ops, etc.) to generate the exact structured query (such as a SQL query) that is executed against the BI platform. The authors propose an automatic bootstrapping mechanism that uses the ontology to generate the intents, their training examples, entities and inputs for building the dialogue to automate the population of the conversational interface artifacts in a domain agnostic way. This provides a repeatable process to quickly build a conversational BI systems for different domains. The system can be further refined using user feedback and inputs from Subject Matter Experts (SMEs).

**Summary**

In this section, we describe conversational interfaces to data which are rapidly gaining popularity as an intuitive technology to democratize access to data. We look at the salient differences in the natural language understanding component of the conversational interfaces compared to text-to-SQL systems. We introduce the concept of conversational semantic parsing that caters to the two-way nature of conversation and provide examples of several state-of-the-art systems that use deep learning approaches for the same. We also introduce dialogue management and its constituent components such as dialogue state tracking as a mechanism for tracking the current state of data exploration given the prior history of queries issued by the user in a data analysis session. We discuss techniques for structured query generation and natural language response generation along with examples of several works in the area that use different techniques to implement the required functionality. Finally, we describe conversational business intelligence as a use case for conversational interfaces to data tailored for BI applications.

# 5

---

# Benchmarks and Evaluation Techniques

---

Evaluating natural language interfaces to data is a non-trivial task (Kaufmann and Bernstein, 2010; Asakura *et al.*, 2018). With the current abundance of solutions that target this problem, a systematic evaluation of existing approaches becomes more and more a necessity. The first steps towards this goal, WikiSQL (Zhong *et al.*, 2017) and Spider (Yu *et al.*, 2018b), have been very well-received by the community, focusing mostly on the learning-based approaches. Recent efforts also focus on providing data sets for evaluating multi-turn and conversational interfaces to data (Yu *et al.*, 2019b; Yu *et al.*, 2019a).

In this section, we present some popular benchmarks for natural language interfaces to data, roughly in chronological order. For more details and more existing benchmarks, we refer the reader to CBench (Orogat *et al.*, 2021) and THOR (Gkini *et al.*, 2021).

## 5.1 WikiTableQuestions

WikiTableQuestions (Pasupat and Liang, 2015) is a benchmark for question answering on semi-structured HTML tables. The data set[1]

---

[1]https://github.com/ppasupat/WikiTableQuestions

contains 2,108 tables from a large variety of topics and 22,033 natural language questions with different complexity. Each question comes with a table from Wikipedia. Given the question and the table, the task is to answer the question based on the table. The whole data set is divided into 14,152 training examples with natural language questions, the table used to answer the question and the answer. There are 4,344 test examples where the table to answer these questions is not previously seen in the training examples. Additionally, the data set provides 3,537 examples targeted as the development data, where the tables to answer the questions are seen in the training data.

## 5.2   WikiSQL

WikiSQL, released along with Seq2SQL (Zhong *et al.*, 2017) covered in Section 3.2, is one of the earliest and most popular benchmarks in the field, containing 80,654 pairs of natural language questions and SQL queries which are manually annotated (via crowd-sourcing, using Amazon Mechanical Turk) and distributed across 24,241 Wikipedia tables. As in WikiTableQuestions, each example in WikiSQL consists of a natural language query, a table and a SQL query corresponding to the natural language query. The examples are randomly split into train, dev, and test sets, making sure that each table appears in exactly one split. In addition to the examples, WikiSQL also provides a corresponding database and query execution engine. The large volume of data enables machine learning based systems to train their model. WikiSQL maintains a leaderboard on its github page[2], reporting the top-ranking supervised and weakly supervised systems.

## 5.3   Spider

Spider (Yu *et al.*, 2018b) is probably the most popular data set for testing the accuracy of text-to-SQL tasks across domains. It consists of 10,181 natural-language questions and 5,693 distinct SQL queries across 200 database schemas with instances, each with multiple tables, covering 138 different domains. Those queries cover a wide spectrum of complex SQL queries, involving joining and nested queries. As shown in a recent study (Sen *et al.*, 2020), as well as in the constantly updated leaderboard chart on Spider's website[3], the variety of the domains as well as the queries in SPIDER present a non-trivial challenge to the accuracy of existing systems, with the highest-scoring ML-based system of the leaderboard (at the time of writing this monograph) reporting an accuracy of 75.1%.

---

[2]https://github.com/salesforce/WikiSQL

## 5.4 SParC

Semantic Parsing in Context (SParC) benchmark (Yu *et al.*, 2019b) is a context-dependent, multi-turn version of the Spider data set. It consists of over 4,000 coherent question sequences[4], obtained from user interactions with 200 complex databases over 138 domains (as in Spider). In order to ensure that each question sequence is relevant in subject and this way, construct meaningful queries, SParC uses questions from the Spider dataset as the thematic guidance. Each question sequence is based on a question in Spider and consists of inter-related questions to obtain the information demanded by different goals. The generated questions are then translated to SQL queries. To ensure correctness, all SQL queries were executed on the same RDBMS (Sqlite) and in order to make the evaluation more robust, the same annotation protocol as the one followed in Spider was adopted, such that the same SQL query pattern is used when multiple queries, equivalent to each other, could be considered.

## 5.5 CoSQL

CoSQL (Yu *et al.*, 2019a), standing for Conversational text-to-SQL[5] is a dialogue version of the Spider and SParC data sets. It consists of more than 30k turns and 10k annotated SQL queries, obtained from the same databases used in the Spider and SParC, following a multi-turn Wizard-of-Oz setup (Budzianowski *et al.*, 2018). In this setup, crowd-sourced workers act as database users, while students proficient in SQL act as database experts. Each dialogue is a simulation of a real database querying scenario with a user trying to explore a database with a pre-defined goal in mind, while an expert retrieves answers with SQL, clarifying those questions that are ambiguous and labeling other questions as unanswerable.

---

[3]https://yale-lily.github.io/spider
[4]https://yale-lily.github.io/sparc
[5]https://yale-lily.github.io/cosql

## 5.6  LC-QuAD

The Large-Scale Complex Question Answering Dataset (LC-QuAD) 2.0 (Dubey *et al.*, 2019) is a dataset consisting of 30,000 pairs of questions and their corresponding SPARQL queries expressed over Wikidata and DBpedia. The questions are semi-automatically created from 22 unique template queries on Wikidata, which are later "verbalized" (i.e., expressed in natural language) and paraphrased via crowd-sourcing, covering 21,258 entities and 1,310 predicates. As in the case of other benchmarks, the project's website[6] maintains a leaderboard with systems having used this dataset.

## 5.7  FIBEN

FIBEN (Sen *et al.*, 2020) is a benchmark dataset that emulates a financial data warehouse. It consists of 300 natural language queries and their corresponding SQL queries, defined over the FIBEN schema. The schema conforms to a union of two standard finance ontologies: Finance Industry Business Ontology[7] and Finance Report Ontology[8]. It contains information about public companies from a variety of industry sectors, their officers, and financial metrics. The FIBEN schema also contains transactions over holdings and securities provided by public companies, where each transaction is linked to a customer's account (containing securities held, buying and selling of securities).

The benchmark includes a mix of 130 single-SQL-block queries, and 170 nested queries that are further classified into four nested SQL types, as defined in (Kim, 1982). Apart from the query examples, FIBEN's website[9] also contains the necessary files (DDL script and CSV instance data) to load the FIBEN benchmark data on a relational database.

---

[6]http://lc-quad.sda.tech/
[7]https://spec.edmcouncil.org/fibo/
[8]http://xbrl.squarespace.com/financial-report-ontology/

## 5.8 CBench

The Carleton Benchmark (CBench) Suite (Orogat *et al.*, 2021) is a framework for evaluating question-answering systems over a target knowledge graph (e.g., DBpedia, Freebase, Wikidata), by utilizing existing benchmark datasets that mostly target SPARQL queries (e.g., LC-QuAD). Since different benchmarks may target different versions of the same knowledge graph, CBench employs a Benchmark Updater module that queries the latest version of the target knowledge graph and retrieves the latest answers. In addition to evaluating question-answering systems, CBench can be also used to analyze other benchmarks. For example, it counts the number of questions per dataset that fall under a specific category, such as How-questions (i.e., questions starting with the word "how"), Wh-questions (i.e., questions starting with "what", "when", "who", etc), and yes/no questions. It also analyzes the frequency of certain keywords and operators in the queries (e.g., SELECT, DISTINCT, GROUP BY, and FILTER), as well as the syntactic and semantic similarity of the queries (i.e., how similar are the embeddings of words used in the queries). CBench is publicly available[10].

## 5.9 THOR

THOR (Gkini *et al.*, 2021) is a text-to-SQL benchmark consisting of 241 queries from 3 datasets (IMDB, MAS, YELP), and the queries are split into 17 expressivity-based categories (e.g., based on whether they involve joins, nesting, group by, aggregates, negation). In addition to effectiveness evaluation, this benchmarking effort also targets to evaluate the efficiency of systems, involving resource consumption, percentage of CPU utilization, scalability, and number of SQL I/Os. To make the efficiency comparison fair, the authors map each evaluated system to a common, generalized reference architecture. Thor's architecture consists of a Parser, responsible for the semantic tagging task (Section 2.3), Indexes and Knowledge Bases, used for the data modeling (Section 2.1), an Entity Mapper and an Interpretation Generator (Section 3.1), and finally a SQL Translator and Executor for generating and running the SQL queries over the underlying database. The dataset and evaluation code of this benchmark are publicly available[11].

---

[9]https://github.com/IBM/fiben-benchmark
[10]https://github.com/aorogat/CBench

### Summary

In this section, we present nine popular benchmarks for natural language interfaces to data, from the least to the most recent. Arguably, the most popular benchmark at the time of writing this monograph is Spider (Yu *et al.*, 2018b), which contains queries from a big diversity of domains. It also maintains a leader board of best-performing ML systems; a very nice idea to encourage competition and push the state-of-the-art systems to improve. We observe that earlier benchmarks contain questions that are relatively simple (e.g., requiring no or few joins and nested queries), while later benchmarks are taking a deeper dive into query complexity. FIBEN (Sen *et al.*, 2020), for example, focuses particularly on nested queries, while THOR (Gkini *et al.*, 2021) further divides queries into multiple expressivity groups, based on whether they involve nesting, aggregates, negation, etc. Yet, we believe that there is still room for improvement, as we discuss in the next section.

---

[11]https://github.com/athenarc/THOR-Text2SQLBenchmarking

# 6

---

# Open Challenges

---

Despite the recent advances in the area of natural language interfaces, there are still many open research challenges. In addition, for this technology to be adopted and widely used by enterprises, there are still several barriers to overcome. In the following, we discuss these open research challenges and adoption barriers.

**Generalizability and domain adaptation.** Most of the proposed systems work reasonably well for benchmarks like WikiSQL, and the data sets they have been built for, but they do not adapt well to new data sets. There has been an increasing focus on domain adaptability but more work is needed for general purpose solutions that work well for any domain, and do not need lots of specialization or training.

**Complexity of queries.** As natural languages are used for more complex tasks, such as data analytics, there is a need for NLIs to produce more complex SQL queries, involving multiple sub-queries. First, detecting whether a natural language query requires to be translated to a nested structured (SQL) query is non-trivial due to non-obvious linguistic patterns and inherent ambiguities in the natural language queries. Second, building a nested query requires identifying proper sub-queries and figuring out the correct conditions to join or combine the sub-queries to produce the correct query results. Hence, building NLIs that can handle both the user utterance complexity as well as can produce complex queries is still an open challenge that requires further research and development.

A promising approach to handle complexity is to break the query into smaller pieces, which can be interpreted into simpler structured queries. Iyyer *et al.* (2017) propose DynSP, a dynamic neural semantic parsing framework for answering sequences of simple related questions. However, they do not actually address the problem of breaking the complex queries into a sequence of simpler queries and use a crowd sourcing approach to manually do this important task. Designing techniques and systems for identifying the complexity of the query, and breaking it into a set of simpler and smaller queries is one of the key challenges towards building effective natural language interfaces to data and their adoption into main stream analytics.

**Hybrid NLQ approaches.** Neither rule-based approaches nor the ML/DL-based text-to-SQL approaches can tackle all the challenges in building natural language interfaces to data. In general, the rule-based approaches provide better accuracy and domain adaptability while the text-to-SQL approaches offer greater flexibility (recall) in terms of the natural language queries, as they are more robust to variations in linguistic patterns. As discussed in Section 3.3, there exist a few hybrid approaches that attempt to leverage the best from both worlds to build effective NLIs. However, these approaches are in their nascent stage and further research is needed to explore these hybrid approaches.

**Training examples.** Text-to-SQL systems and emerging hybrid approaches rely on ML/DL models to interpret user intent. The more variety of queries the system needs to handle, the more training examples the system needs. Generating such training data for ML/DL models for NLI tasks is both time-consuming as well as labor intensive. Conversational analytical interfaces can be built by exploiting common analytics query patterns, and these patterns can also be used to generate training examples (Quamar *et al.*, 2020b).

**Trust and explainability.** It is vital for users to trust the NLI solution that their utterances are interpreted correctly. This is even more pronounced in analytics use cases where the user gets charts and graphs depicting trends as a result. NLG techniques are used in most of these systems to tell the user how their queries are interpreted to build trust.

As part of the broader topic of responsible data management, along with algorithmic fairness and transparency (Stoyanovich *et al.*, 2020; Stoyanovich *et al.*, 2016), the importance of explainability has emerged as a key issue in building natural language interfaces. In particular, deep learning-based natural language interfaces to data often come at the expense of models becoming less interpretable, which may erode trust in these systems. Natural language interfaces need to be integrated with major explainability techniques, such as feature importance, provenance, declarative induction, as well as the most commonly used explainability and visualization techniques, operations used to generate explanations in the NLP literature (Danilevsky *et al.*, 2020), in order to provide reasonable explanations through text generated using NLG techniques. Explanations can be further distinguished into those answering "HOW" and those answering "WHY" (or even "WHY NOT") questions (Preece, 2018). The former type of questions mostly refer to understanding the workings of a system that led to an answer (e.g., a trace of the rules triggered to produce this answer), while the latter may further question the reasoning process behind a provided answer, which leads to a what is also referred to as causal understanding (Holzinger *et al.*, 2019).

**Conversational data analysis.** There are several challenges for building conversational interfaces for data exploration and analysis. These include challenges in conversational semantic parsing, dialog state tracking and dialog modelling using deep learning techniques including pre-trained language models. Each of these building blocks needs to address the challenges of scalability, low latency as well as high accuracy. Training models for these building blocks requires a large amount of conversational training data. These conversational data sets are not readily available and require a lot annotation effort to make them useful. This is further aggravated by issues arising from domain adaptability. Building domain-specific conversational interfaces for exploring data is a major challenge where capturing the domain semantics and incorporating that into the conversational system is non-trivial. This includes understanding the entities of the domain and their relationships, as well as the domain vocabularies and their synonyms.

**Benchmarks.** We believe that the next step to follow up on the benchmarking efforts mentioned in Section 5 is the inclusion of more complex analytical queries in these benchmarks that will enable robust testing of the rule-based approaches and further push the state-of-the-art for text-to-SQL and hybrid approaches as well.

**High precision.** Natural language interfaces can democratize access to data within an enterprise by enabling non-technical users to explore the data easily. However, many enterprise applications require high accuracy, and the current state-of-the-art approaches still cannot achieve desirable levels. For natural interfaces to become widely adopted in the enterprise more research is needed to increase the precision while maintaining high recall for both simple and complex queries.

# 7

# Conclusion

Natural language interface to data has been an active area of research for more than a decade (Özcan *et al.*, 2020; Li and Rafiei, 2017; Affolter *et al.*, 2019; Katsogiannis-Meimarakis and Koutrika, 2021b; Gkini *et al.*, 2021). With recent advances in NLP technologies, not only research in this field has seen a resurgence, but also practical solutions started to appear in commercial products (*Ask Data | Tableau Software* 2021; *Power BI Platform* 2021; *Cognos Assistant* 2021). In this monograph, we review natural language interfaces to data, which include both NLQ systems as well as conversational solutions. We describe various technologies used in natural language interfaces, including semantic tagging, language models, ontologies and semantic indexing techniques.

We describe many solutions that fall into two broad categories: rule-based and text-to-SQL. The rule-based solutions utilize semantic indexes, ontologies and knowledge graphs to identify entities in the user utterance, correlate those entities with the elements in the database, identify a meaningful relationship between them, and create an interpretation of the user intent. Most rule-based systems use templates, grammars, or rules to generate a SQL statement from this interpretation. Text-to-SQL systems, on the other hand, use deep learning techniques, mostly variations of sequence-to-sequence transformers to translate a user utterance into a SQL query in a holistic way. Some systems combine multiple models for semantic tagging and interpretation steps. While the rule-based systems capture the semantics and domain information better than the text-to-SQL systems, the latter is more robust to language variations. There is also an emerging third category of hybrid systems which combine the strengths of rule-based techniques and text-to-SQL systems for more robust and accurate interpretation.

We also discuss conversational systems to data analytics. Gartner (Richardson *et al.*, 2021) predicts that conversational interfaces will replace dashboards in the future, with the consumer-focused, augmented in context, conversational analytics experiences. We describe several challenges in conversational data exploration, including conversational semantic parsing, dialogue management, and structured query generation. By providing context across multiple turns of conversation, these systems support data exploration in a natural way.

With all these research and development activity, it is important to have benchmarks to track the progress of NLI systems and enable the reproduction of results. As such, we also provide an overview of these benchmarks, and discuss their strengths and shortcomings.

Natural language interface is still a developing technology where the accuracy and precision of the solutions need improvement for wider enterprise adaption. Another important impediment is the trust in the interpretations, as well as the explainability of the produced queries. There is also room for improving the complexity of both the input user utterance, as well as the generated structured query. Nevertheless, natural language interfaces democratize access to data for all users, and we expect them to become more commonplace in the near future.